\documentclass[10pt,journal,a4paper]{IEEEtran}
\usepackage{times,amsmath,epsfig,amsfonts,amssymb,graphicx,url,cite,subfigure,colortbl,bm,upgreek}
\usepackage{color,soul}

\renewcommand{\baselinestretch}{1}
\DeclareRobustCommand*{\IEEEauthorrefmark}[1]{\raisebox{0pt}[0pt][0pt]{\textsuperscript{\footnotesize #1}}}

\begin{document}
	\title{Antenna Systems for Wireless Capsule Endoscope: Design, Analysis and Experimental Validation}
	
	\author{Md. Suzan Miah, Ahsan Noor khan, Clemens Icheln, Katsuyuki Haneda,~\IEEEmembership{Member, IEEE}, and Ken-ichi Takizawa% <-this % stops a space                                    % ...
		%\\
		\thanks{Manuscript received Month Day, Year; revised Month, Day, Year; accepted Month, Day, Year. Date of publication Month, Day, Year; date of current version Month, Day, Year.}
		\thanks{M. S. Miah, K. Haneda and C. Icheln are with the Department of Electronics and Nanoengineering, Aalto University-School of Electrical Engineering, Espoo FI-00076, Finland. (e-mail: md.miah@aalto.fi).}% <-this % stops a space
		\thanks{A.N. Khan was with the Department of Electronics and Nanoengineering, Aalto University-School of Electrical Engineering, Espoo FI-00076, Finland. (e-mail: ahsannoorkhan@gmail.com).}
		\thanks{K. Takizawa is with the National Institute of Information and Communications Technology (NICT), 3-4 Hikarino-oka, Yokosuka, Japan 2390847. (e-mail: takizawa@nict.go.jp).}
	}
	\vspace{-20pt}
	
	\markboth{IEEE Transactions ON ANTENNAS AND PROPAGATION - For review}%
	{Suzan \MakeLowercase{\textit{et al.}}: Antenna systems for Wireless Capsule Endoscopy}
	
	\maketitle

\begin{abstract}
Wireless capsule endoscopy (WCE) systems are used to capture images of the human digestive tract for medical applications. The antenna is one of the most important components in a WCE system. In this paper, we provide novel small antenna solutions for a WCE system operating at the $433$ MHz ISM band. The in-body capsule transmitter uses an ultrawideband outer-wall conformal loop antenna, whereas the on-body receiver uses a printed monopole antenna with a partial ground plane. A colon-equivalent tissue phantom and CST Gustav voxel human body model were used for the numerical studies of the capsule antenna. The simulation results in the colon-tissue phantom were validated through \textit{in-vitro} measurements using a liquid phantom. According to the phantom simulations, the capsule antenna has $-10$ dB impedance matching from $309$ to $1104$ MHz. The ultrawideband characteristic enables the capsule antenna to tolerate the detuning effects due to electronic modules in the capsule and due to the proximity of various different tissues in gastrointestinal tracts. The on-body antenna was numerically evaluated on the colon-tissue phantom and the CST Gustav voxel human body model, followed by \textit{in-vitro} and \textit{ex-vivo} measurements for validation. The on-body antenna exceeds $-10$ dB impedance matching from $390$ MHz to $500$ MHz both in simulations and measurements. Finally, this paper reports numerical and experimental studies of the path loss for the radio link between an in-body capsule transmitter and an on-body receiver using our antenna solutions. The path loss both in simulations and measurements is less than $50$ dB for any capsule orientation and location. 
		
\end{abstract}
	{\smallskip \keywords Conformal antenna, in- to on-body propagation, wireless capsule endoscope.}
	\IEEEpeerreviewmaketitle
	
	\section{Introduction}
	Wireless capsule endoscopy (WCE) is used to record images of the digestive tract for medical applications \cite{G_Iddean, Kiourti_review}. One of the use cases of capsule endoscopy is to examine areas of the small intestine that cannot be seen by traditional types of endoscopy, such as colonoscopy and esophagogastroduodenoscopy. Traditional techniques are painful and time-consuming, whereas WCE is non-invasive and painless. The patient swallows a small capsule with a tiny camera embedded into it. The capsule moves through the gastrointestinal (GI) tract and takes images, which are transmitted to a receiver unit outside the body of the patient. A physician interprets these images either in real-time or offline. 
	
The antennas of the capsule transmitter and the on-body receiver are important components in the WCE system.  Recent research activities on capsule antennas have shown embedded \cite{E_hatmi,E_FMerliTAP,E_Huang_letter,E_SI_kwak2,E_SiKwak, Em_Lee_Yoo1,E_lee_dual_spiral,E_sang_lee_conical,E_sang_lee_Tbio,E_sanglee_189} and conformal structures \cite{Xiang_Senior_2_4,c_denys_TAP2017,c_JFaerber_Tcir_2017,Kiourti_P,c_xiang_L,pathloss_3_Yann_mahe,arefin,c_lijie_AWPL,c_KKwon,c_rupom_das_TAP2017,Izdebski_TAP,c_rajagopalan,Sumin_Kim_TAP_L,Jing_Lim,Rula_huang} as promising antenna types. Embedded antennas are placed inside the capsule cavity. The conformal structure utilizes only the surface of the capsule module and leaves the interior for other components, allowing the most effective use of available surface area of capsule and the antenna to be larger for better radiation performance\cite{Sumin_Kim_TAP_L}. As to the operational radio frequency of a WCE system, \cite{400_600MHz} proposes the frequency band $400-600$ MHz because of a minimum propagation loss in human body tissues. Similarly \cite{Baser_jpier_pathloss} shows that frequencies between $400$ and $500$ MHz are most suitable. In order for the capsule antennas to be insensitive to surrounding human tissue, magnetic antennas are preferred over electric antennas \cite{E_hatmi,Roopnariane}; loop antenna is one of the typical magnetic antennas. Therefore, a conformal loop antenna operating at frequencies between $400$ and $600$ MHz is chosen in this study. 
	
Several designs of a conformal antenna for WCE systems operating at frequencies upto $1400$ MHz have been reported in literature \cite{c_denys_TAP2017,c_JFaerber_Tcir_2017,Kiourti_P,c_xiang_L,pathloss_3_Yann_mahe,arefin,c_lijie_AWPL,c_KKwon,c_rupom_das_TAP2017,Izdebski_TAP,c_rajagopalan,Sumin_Kim_TAP_L,Jing_Lim,Rula_huang}. The antenna needs to fit into a small capsule, and the bandwidth requires to be wide to overcome the detuning effects due to varying tissue properties through the GI tract as well as to realize higher data rate. The conformal antennas in \cite{c_denys_TAP2017,c_JFaerber_Tcir_2017,Kiourti_P,c_xiang_L,pathloss_3_Yann_mahe} report bandwidths upto $53$ MHz, whereas in \cite{arefin,c_lijie_AWPL,c_KKwon,c_rupom_das_TAP2017} it is upto $185$ MHz. The antennas in \cite{Izdebski_TAP,c_rajagopalan} operate at $1400$ MHz with $200$ MHz bandwidth. An outer-wall loop antenna is presented in \cite{Rula_huang} reports a bandwidth of $785$ MHz, while a conformal printed inverse-F (PIFA) antenna with the bandwidth of $562$ MHz is presented in \cite{Jing_Lim}, but, no experimental validation has been reported. The outer-wall loop antenna in \cite{Sumin_Kim_TAP_L} works at the center frequency of $500$ MHz with $260$ MHz measured bandwidth. Moreover, radiation characteristics of the ingestible antenna depend on position, surrounding tissues and orientation of the antenna \cite{Position_orientation_L_Xu,Dielectric_effect_L_Xu,Position_orientation_Chirwa,pathloss_4_khaleghi_IN_ON}. However, none of the previous works has rigorously studied them.

Finally, the characterization of the radio link between in-body capsule transmitter and on-body receiver is an important step in the development of the WCE system. Many studies exist on radio propagation of such a link, e.g., \cite{pathloss_1_anzai_IN_ON, pathloss_2_shie, pathloss_3_Yann_mahe, pathloss_4_khaleghi_IN_ON, pathloss_5_Raul_IN_ON, pathloss_6_Karen_lopez, pathloss_7_liu, pathloss_8_pal_IN_ON, pathloss_9_garcia_padro_IN_ON, pathloss_10_kasun,pathloss_Aloainy}. However, most of the literature presents in-to-off body propagation, where the outside-body antennas are not in contact with the body \cite{pathloss_2_shie, pathloss_3_Yann_mahe, pathloss_5_Raul_IN_ON, pathloss_6_Karen_lopez, pathloss_7_liu, pathloss_10_kasun, Ding_jpier}, even though it is very advantageous to place the antenna near the abdomen. The models discussed in \cite{pathloss_1_anzai_IN_ON, pathloss_5_Raul_IN_ON, pathloss_8_pal_IN_ON, pathloss_10_kasun} cover the path loss between in-body and on-body antennas, but for frequencies above $1$ GHz. The in- to on-body propagation channel models for ingestible medical applications at frequencies of $402$ MHz, $868$ MHz, and $2400$ MHz have been discussed in \cite{pathloss_Aloainy}. However, the effects of changing orientation of implant on the path loss are not investigated. In the realization of a reliable WCE system, the on-body antenna requires to work efficiently in the proximity of human body. Some literatures are available on the design of on-body antennas that include the human body effect \cite{Rec_park, pathloss_1_anzai_IN_ON, pathloss_4_khaleghi_IN_ON, pathloss_5_Raul_IN_ON, pathloss_8_pal_IN_ON, pathloss_10_kasun}, but their focus is on higher frequencies than $500$ MHz. For below-1~GHz on-body antennas, \cite{pathloss_4_khaleghi_IN_ON} presents a spiral helix antenna at $600$ MHz; \cite{Rec_park} also shows such solutions, but the dielectric properties of the tissue of the body model in their study are not specified. In \cite{Dual_band_repeater_kiourti, Dual_band_repeater_L_Xu}, dual-band on-body repeater antennas at MedRadio ($401-406$ MHz) and $2400$ MHz ISM band have been presented. To the authors' best knowledge, there is no literature reporting on-body receiver antenna of a WCE system operating at a radio frequency of $433$ MHz industrial, scientific and medical (ISM) band. Subsequently, there are no studies on path loss for a WCE system operating at $433$~MHz. In summary, the contributions of this paper are therefore threefold: 
\vspace{-10pt}   

\begin{enumerate}
\item[a)] We propose a novel ultrawideband conformal loop antenna attached on the outer-wall of a capsule module operating at $433$ MHz ISM band. The designed antenna operation is experimentally verified with a prototype and measurements in a colon-equivalent liquid phantom. The $-10$ dB impedance matching of the antenna is between $309$~MHz and $1104$~MHz, outperforming the existing solutions. It was also demonstrated that the antenna is robust against changes of surrounding environments such as other components in the capsule, different tissues in the GI tract, different locations and varying orientation inside the body. Note that the proposed capsule antenna is an extension of the authors' own work \cite{Suzan_eucap}. The technical advances of the present paper compared to \cite{Suzan_eucap} include i) improved bandwidth by optimizing antenna structure and dimensions, ii) numerically evaluating resonance and radiation performance using a single-layer colon phantom as well as a CST Gustav voxel human body model, iii) conforming the antenna on a realistic capsule module and iv) a study of the specific absorption rate (SAR).     
\item[b)] We propose a novel monopole antenna with a partial ground plane for the on-body receiver unit. The antenna has $-10$~dB impedance matching from $390-500$~MHz. Comparisons of \textit{in-vitro} and \textit{ex-vivo} on-body antenna measurements with numerical simulations on a colon tissue phantom and a CST Gustav voxel human body model show sufficient agreement of the matching characteristics.  
\item[c)] The in- to on-body link path loss is evaluated for the developed antennas. \textit{In-vitro} measurements for various locations and orientations of the capsule antenna show reasonable agreement with corresponding numerical simulations, demonstrating the validity of the entire approach of our antenna design and path loss simulations. The path loss is less than $50$~dB regardless of the capsule locations and orientations within the body.        
	\end{enumerate}
The remainder of this paper is organized as follows. Section~II illustrates the proposed capsule antenna configuration and simulations, while on-body antenna structure and simulations are described in Section~III. Section~IV addresses the experimental validation of the capsule and on-body antenna, and path loss. Finally, Section~V concludes this paper.
	
\section{Capsule Antenna Design and Simulations}
\subsection{Antenna Structure}
	\label{sec:capsule_antenna_structure}
\begin{figure}[t!]
		\begin{center}
			\subfigure[]{\resizebox{.6\hsize}{!}{\includegraphics[scale=1.2]{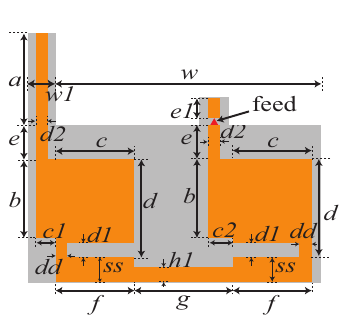}}
				\label{fig:before_wrap}}
			\subfigure[]{\includegraphics[scale=1.4]{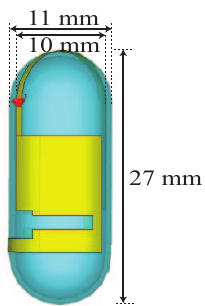}
				\label{fig:after_wrap}}
			\caption{Capsule antenna (a) before, and (b) after placing around the capsule.}
			\vspace{-10pt}
			\label{fig:antenna_structure}
		\end{center}
	\end{figure}

 The proposed antenna is a loop antenna patterned on a $100$ $\mu$m thick flexible substrate Preperm 255, which allows bending and wrapping around the capsule. Relative permittivity,  $\varepsilon_{\rm r}$, and loss tangent, $\tan\delta$, of the substrate are $2.55$ and $5.0 \times 10^{-4}$, respectively. Copper of $19~{\rm \mu m}$ thickness is used as a conductor material on the substrate. The proposed antenna before and after wrapping it around the capsule is shown in Fig.~\ref{fig:antenna_structure}. The antenna utilizes the outer-wall of the cylinder and one dome of the capsule module, whereas the other dome remains free for the camera and other optical components. The capsule module is made of polystyrene with $\varepsilon_{\rm r} =2.6$ and $\tan \delta =0.05$ at $1$ GHz. The thickness of the capsules' wall, diameter, and length of the capsule are $0.5$~mm, $11$ mm and $27$ mm, respectively. The antenna does not have a ground plane in order to avoid strong mirror current, which reduces antenna efficiency. Besides other parameters of the antenna change compared \cite{Suzan_eucap}, an additional slot has been introduced in each loop arm as shown in Fig.~\ref{fig:before_wrap} to enhance bandwidth. The dimensions of the loop have been optimized to resonate at $433$~MHz as summarized in Table~\ref{table:antenna_dimensions}. The feeding point of the antenna in the simulations is indicated by a red triangle in Fig.~\ref{fig:after_wrap}.
  %The diameter and length of the capsule module are $11$ mm and $27$ mm, respectively, whereas the thickness of the capsules' wall is $0.5$~mm.
	\vspace{-10pt}	
	
	\begin{table}
		\begin{center}
			{\caption{Optimized dimensions of the capsule antenna structure in mm} 
				\label{table:antenna_dimensions}}
			\begin{tabular}{m{.2cm}m{.1cm}m{.1cm}m{.1cm}m{.1cm}m{.1cm}m{.1cm}m{.1cm}m{.1cm}m{.1cm}m{.1cm}m{.1cm}m{.1cm}m{.1cm}m{.1cm}m{.2cm}m{.1cm}} \hline
				$a$  & $b$  & $c$ & $d$ & $e$ & $f$ & $g$ & $c1$& $c2$ & $d1$ & $d2$ & $e1$ & $dd$ & $ss$& $h1$& $w$& $w1$ \\
				
				\hline
				14.7 & 8 & 8 & 10 & 3.5 & 8 & 10 & 2 & 2.5&1.5& 1.2 & 2 & 1.5 & 2.5 & 1.5 & 31.4&3.2\\
				
				\hline
			\end{tabular}
		\end{center}
		\vspace{-15pt}
	\end{table}
	
	\subsection{Capsule Implementation and Operating Environments}
	\label{sec:capsule_implementation_and_operating_environments}
\begin{figure}[t!]
		\begin{center}
			\resizebox{1\hsize}{!}{\includegraphics[scale=.8]{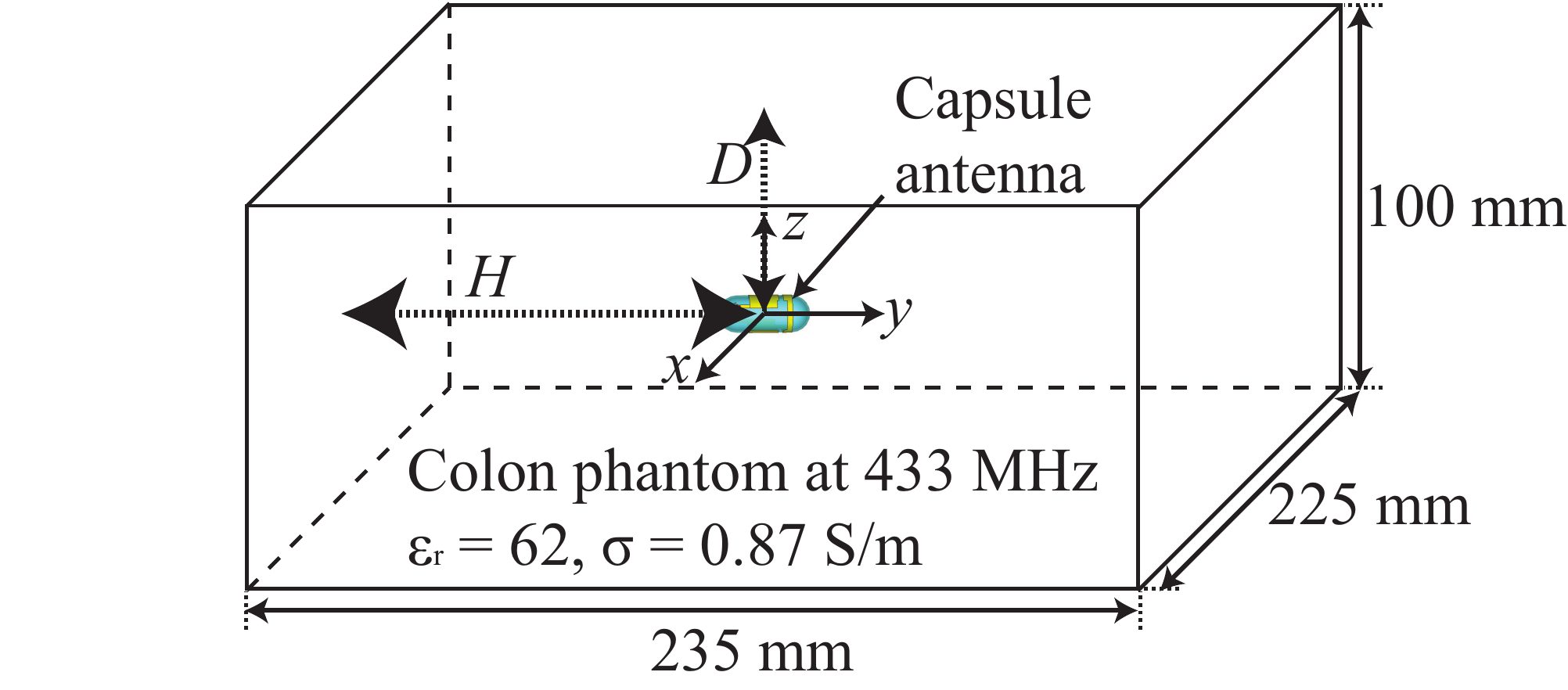}}
			\caption{The capsule endoscope antenna in the single tissue phantom model, where the dielectric properties of the colon-tissue have been used.}
			\vspace{-15pt}
			\label{fig:sim_arrangement}
		\end{center}
	\end{figure}

	A single-layer colon-tissue phantom model with dimensions of $235$ mm $\times$ $225$ mm $\times$ $100$ mm was used for the proposed antenna design and optimization. The antenna arrangement in Fig.~\ref{fig:after_wrap} was implanted at the center of a  colon tissue phantom as visualized in Fig.~\ref{fig:sim_arrangement}. The colon tissue is more relevant than muscle tissue for our simulations as it is one of the organs of the human GI tract, we preferred to use colon-tissue phantom instead of widely used muscle tissue. The dielectric properties of the colon tissue are frequency dependent, and these at $433$ MHz are listed in Table~\ref{table:different_tissue}. We considered the frequency dependency when simulating antenna matching across wide bandwidth centered at $433$ MHz. The CST Studio Suite 2017 was used for the simulations.

	%We study numerically the changes of antenna matching for various implementations of the capsule and its operations, including different orientations and locations of the capsule. Different lossy tissues the capsule would pass through in practice were considered. A single-layer \textcolor{red}{colon equivalent} tissue phantom model is used for the antenna design and also in the prototype measurements. \textcolor{red}{Since, during the test period capsule passes through the colon, which is one of the organs in the human GI tract, we prefer to use colon-tissue instead of popular muscle tissue phantom}. The CST Studio Suite 2016 is used for the antenna \textcolor{red}{optimization}. The simulation setup is shown in Fig.~\ref{fig:sim_arrangement}. Length, width and height of the rectangular phantom are $235$ mm, $220$ mm and $120$ mm, respectively. As a capsule travels through the entire GI tract, such as the esophagus, stomach, small intestine and colon, it experiences a significant change of relative permittivity and conductivity. The dielectric properties of those organs at $433$ MHz are presented in Table~\ref{table:different_tissue} ~\cite{C_Gabriel}. 
\subsubsection{A Reference Case}

\begin{figure}[t!]
	\begin{center}
		\resizebox{0.80\hsize}{!}{\includegraphics[scale=.45]{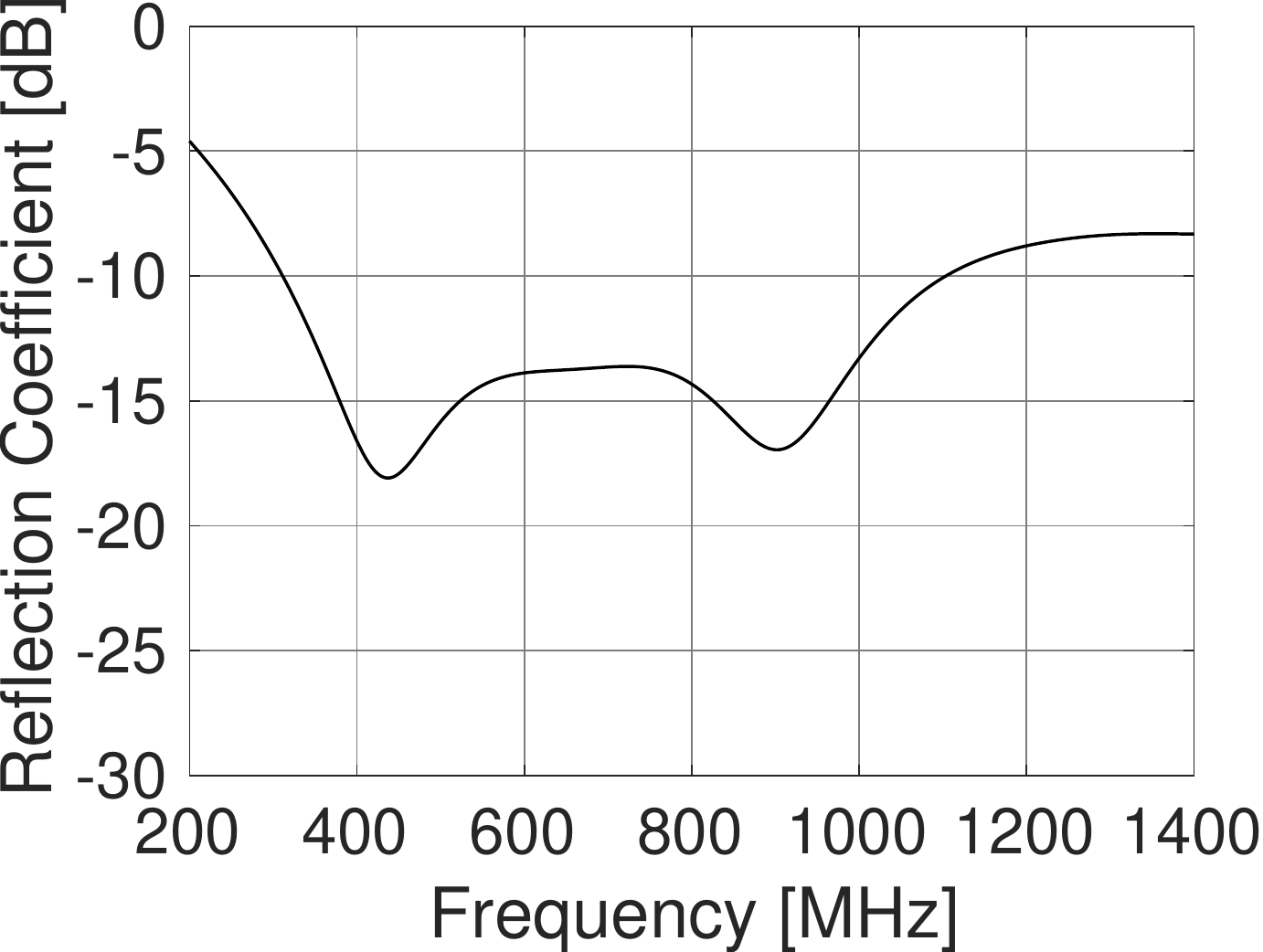}}
		\caption{Simulated reflection coefficient of the $Y$-orientated capsule antenna at the center of the colon-tissue phantom.}
		\vspace{-10pt}
		\label{fig:sim_in_body_antenna_at_center}
	\end{center}
\end{figure}

First, the capsule antenna without biocompatible layer and electronic components in the capsule was placed in the center of the colon-tissue phantom as shown in Fig.~\ref{fig:sim_arrangement}; $Y$-oriented means that the longest dimension of the capsule is aligned with the $Y$-axis. The distance between the center of the capsule to the top of the phantom and to the side wall of the phantom are $50$ mm and $117.5$ mm, respectively. Fig.~\ref{fig:sim_in_body_antenna_at_center} shows the simulated magnitude of a reflection coefficient, $|S_{11}|$. The antenna exhibits double resonances, at $438$~MHz and $905$ MHz. The $-10$ dB impedance matching is achieved across $309$~MHz to $1104$~MHz, which covers the entire band of interest. The peak realized gain of the antenna is $-35$ dBi at $433$ MHz. In the following studies, this serves as the baseline, and the above-mentioned settings and the same colon-tissue model are used unless otherwise stated.

%Fig.~\ref{fig:sim_in_body_antenna_at_center} shows the simulated magnitude of reflection coefficient, $|S_{11}|$, when the capsule antenna without biocompatible layer is placed in the center of the \textcolor{red}{colon-tissue} phantom \textcolor{red}{as shown in Fig.~\ref{fig:sim_arrangement}}; $Y$-oriented means that the longest dimension of the capsule is aligned along the $Y$-axis. The antenna resonates at $420$~MHz with an $|S_{11}|$ of about $-19$ dB. The proposed capsule antenna shows $-10$ dB impedance matching from $300$~MHz to $875$~MHz, which covers the entire band of interest from $400$ to $600$ MHz. In the following studies, this serves as a reference case, and the above mentioned settings \textcolor{red}{and colon-tissue model} are used unless otherwise stated. 
%Parametric studies play a significant role in the design, as well as in optimization of an antenna to achieve good matching at the desired frequency. Based on the optimized geometry parameters, we studied the significant capsule antenna parameters for tuning and bandwidth enhancement. For simplicity, the biocompatible material layer and the electronic components were not considered in the initial antenna design. Later, we will study the effect of biocompatible layer and electronics in Section~\ref{sec:effect_of_biocompatible_layer} and Section~\ref{sec:effect_of_electronics}, respectively.
\subsubsection{Parametric Studies of the Proposed Capsule Antenna}
\label{sec:parametric analysis of the capsule antenna}
\begin{figure}[t!]
	\begin{center}
		\resizebox{0.8\hsize}{!}{\includegraphics[scale=.45]{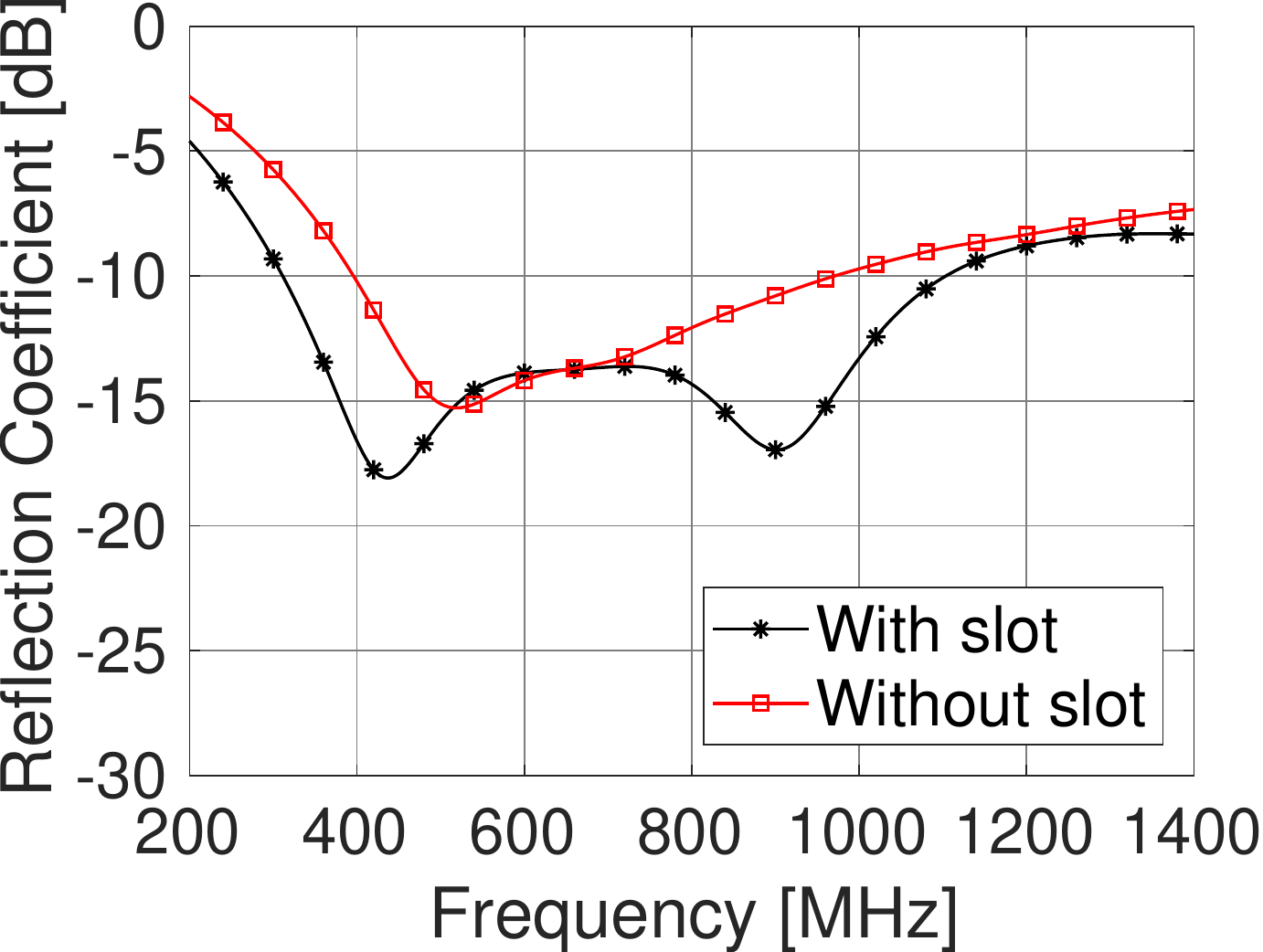}}
		\caption{Effects of the slots on simulated reflection coefficient of the $Y$-orientated capsule antenna at the center of the colon-tissue phantom.}
		\vspace{-10pt}
		\label{fig:TAP1_sim_with_and_with_slot}
	\end{center}
\end{figure}

 We consider the $Y$-oriented capsule antenna at the center of the phantom (see Fig.~\ref{fig:sim_arrangement}). For the proposed antenna, the most important antenna parameters were the thickness and position of the slots, i.e., $d1$ and $ss$ in Fig.~\ref{fig:before_wrap}. We begin investigating the effect of the slots on the antenna matching. Fig.~\ref{fig:TAP1_sim_with_and_with_slot} shows that the simulated $|S_{11}|$ with and without slots in the antenna geometry. We can see that the antenna without slot resonates at around $500$ MHz with a $-10$ dB impedance bandwidth of $572$ MHz. After introducing the slots, the bandwidth increased to $795$ MHz.
    
\begin{figure}[t!]
	\begin{center}
		\resizebox{0.80\hsize}{!}{\includegraphics[scale=.45]{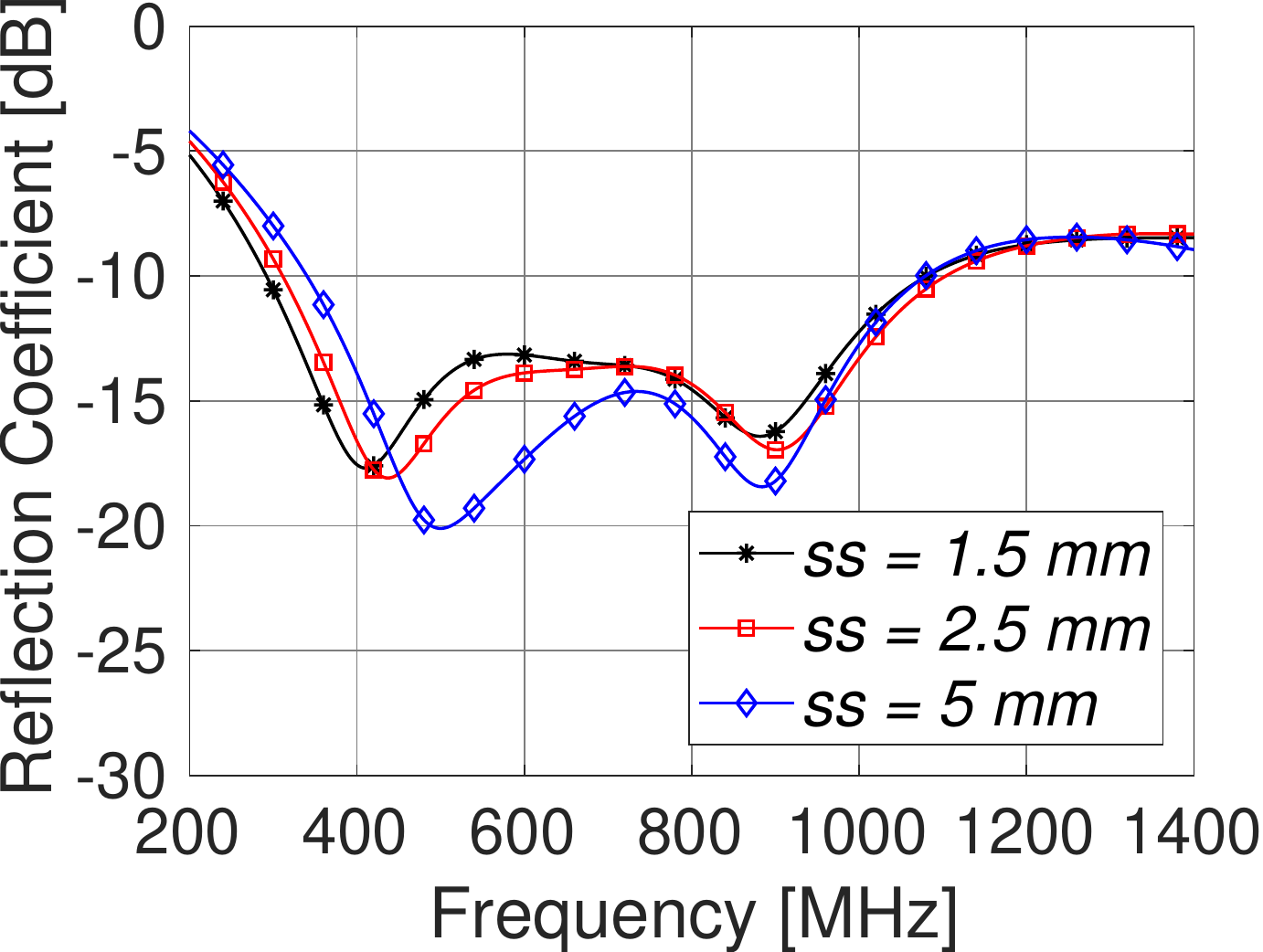}}
		\caption{Effects of the position of the slots on simulated reflection coefficient of the $Y$-orientated capsule antenna at the center of the colon-tissue phantom.}
		\vspace{-20pt}
		\label{fig:TAP1_sim_slot_move_ss}
	\end{center}
\end{figure}

\begin{figure}[t!]
	\begin{center}
		\resizebox{0.80\hsize}{!}{\includegraphics[scale=.45]{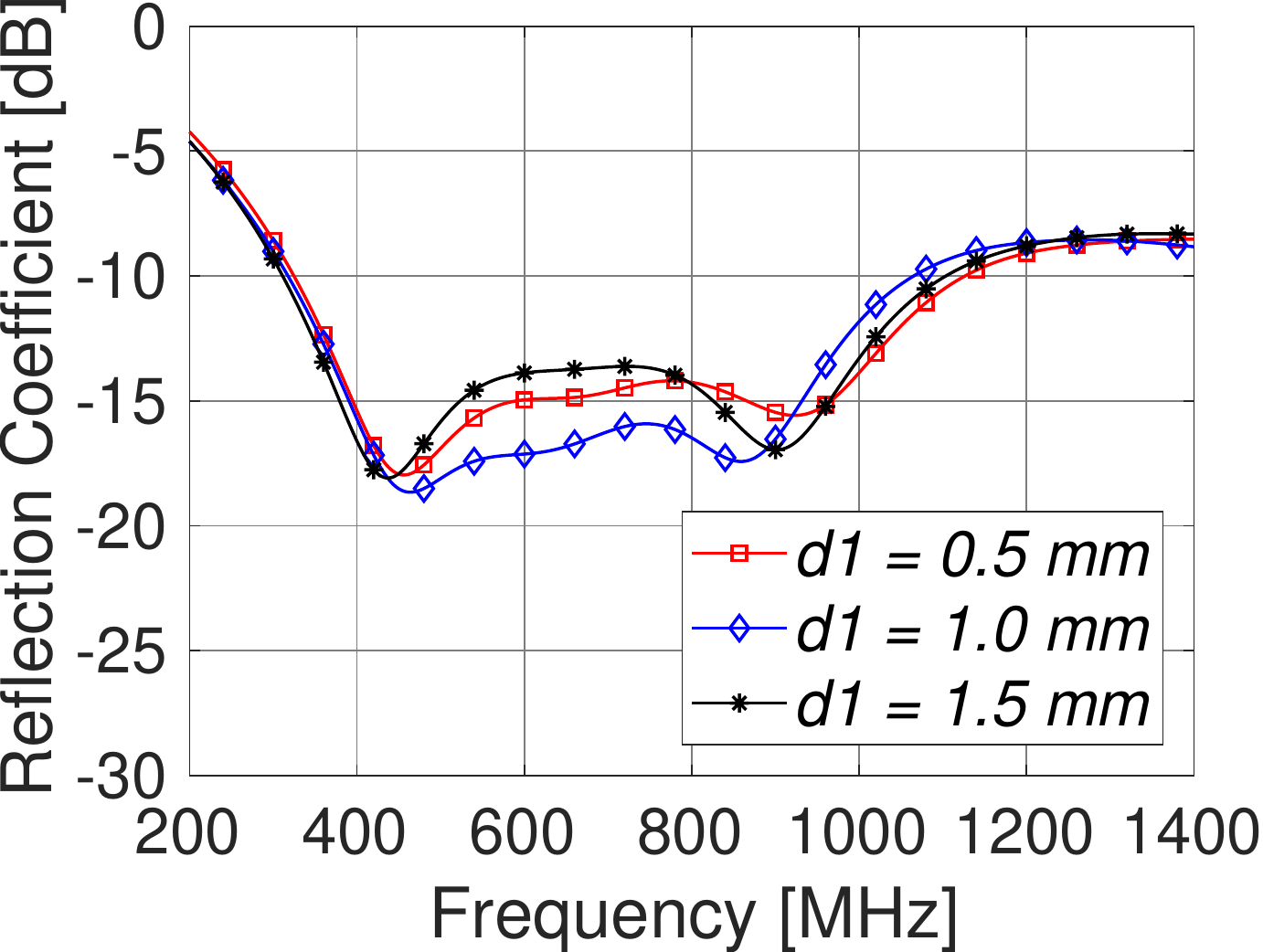}}
		\caption{Effects of the thickness of the slots on simulated reflection coefficient of the $Y$-orientated capsule antenna at the center of the colon-tissue phantom.}
		\vspace{-20pt}
		\label{fig:TAP1_sim_slot_thickness}
	\end{center}
\end{figure}

Fig.~\ref{fig:TAP1_sim_slot_move_ss} clearly shows that the first resonance at around $500$ MHz appears as we increase $ss$ to $5$ mm, where $ss$ represents the distance of the slot from the bottom of the antenna as indicate in Fig.~\ref{fig:before_wrap}. The resonance is shifted downwards as we decrease $ss$. Results show $57$ MHz improvement in the impedance bandwidth as we decrease $ss$ from $5$ to $2.5$ mm. Fig.~\ref{fig:TAP1_sim_slot_thickness} shows the effect of the thickness of the slot dimension $d1$ on the matching. The first resonance appears at around $460$ MHz for $d1 = 0.5$ mm and shifts to $438$ MHz as $d1$ increases to $1.5$ mm. For our goal to achieve a resonance at $433$ MHz with maximum bandwidth, we choose $ss$ and $d1$ to be $2.5$ and $1.5$ mm, respectively.  

\subsubsection{Effects of Capsule Orientations}
\label{sec:implant_orientations}
As it is impossible to control the orientation of the capsule during endoscope operation, its orientation is considered random. We numerically evaluate the changes of antenna matching for three different orientations of the capsule at the center of the colon-tissue phantom in Fig.~\ref{fig:sim_arrangement}. Antenna matching for $X$-, $Y$- and $45^\circ$ slanted in $YZ$-oriented capsule is shown in Fig.~\ref{fig:sim_in_body_implant_orientations}. The results demonstrate that the capsule orientation does not have a significant impact on the antenna resonance, as can be expected inside a large homogeneous phantom. The peak realized gain at $433$ MHz of the $X$-oriented, $Y$-oriented and $45^\circ$ slanted (in $YZ$-plane) capsule antenna is $-34.5$, $-35$ and $-32$ dBi, respectively. The variation of the peak gain across different orientation is not significant. 
\subsubsection{Effects of Biocompatible Layer}
\label{sec:effect_of_biocompatible_layer}

\begin{figure}[t!]
		\begin{center}
			\subfigure[]{\includegraphics[scale=.3] {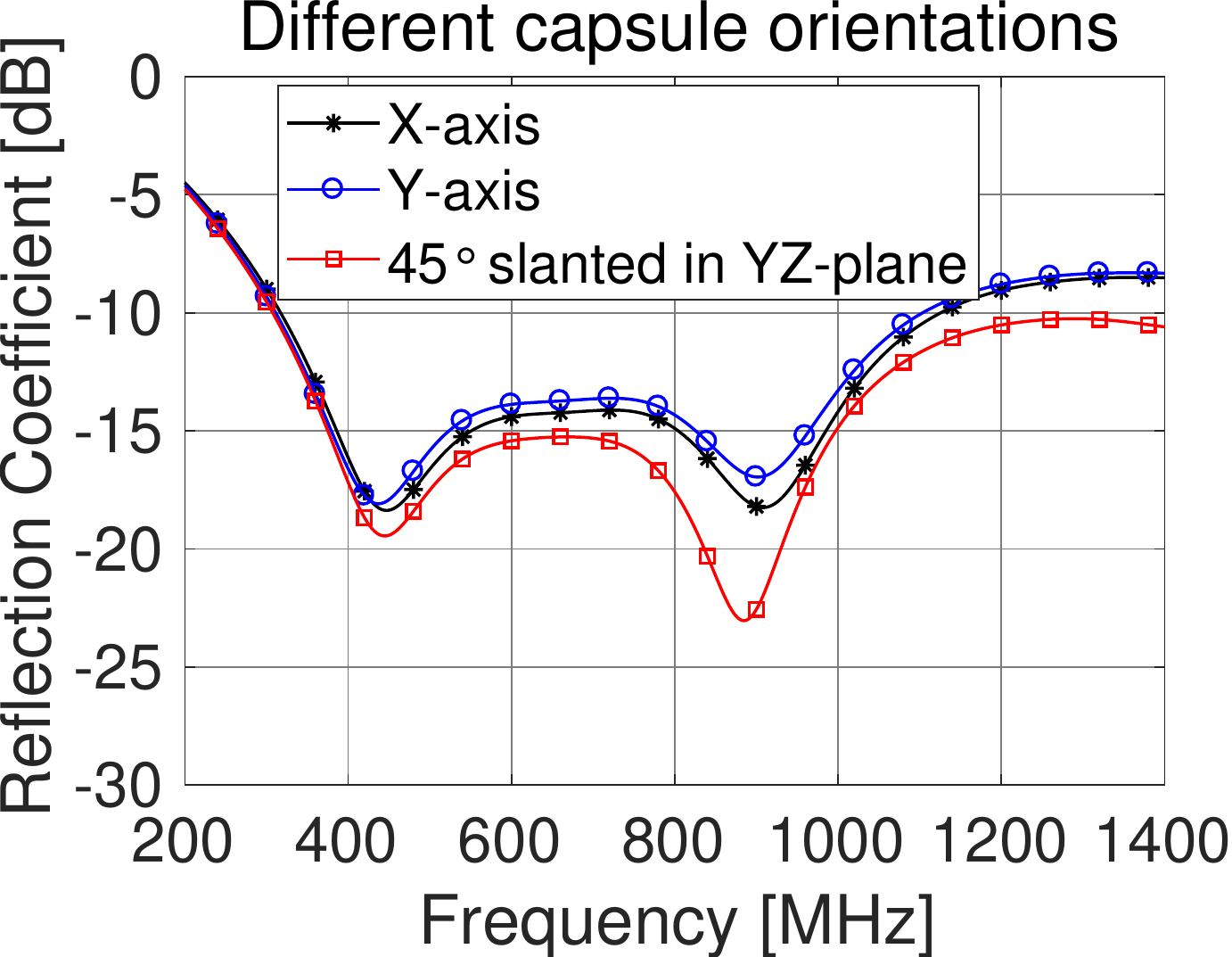}
				\label{fig:sim_in_body_implant_orientations}} 
			\subfigure[]{\includegraphics[scale=.3] {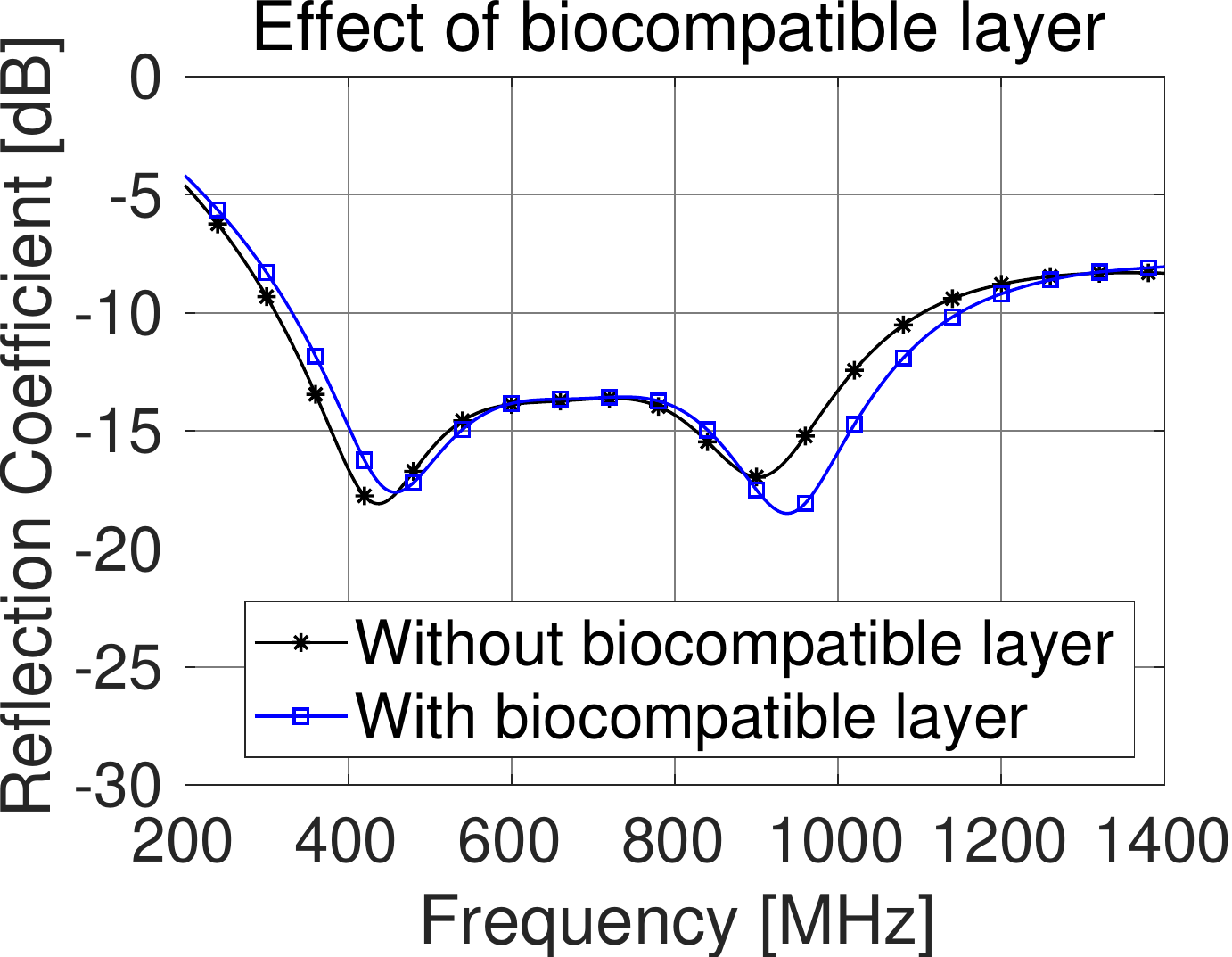}
				\label{fig:sim_with_and_with_bio_layer}}
			\subfigure[]{\includegraphics[scale=.3]{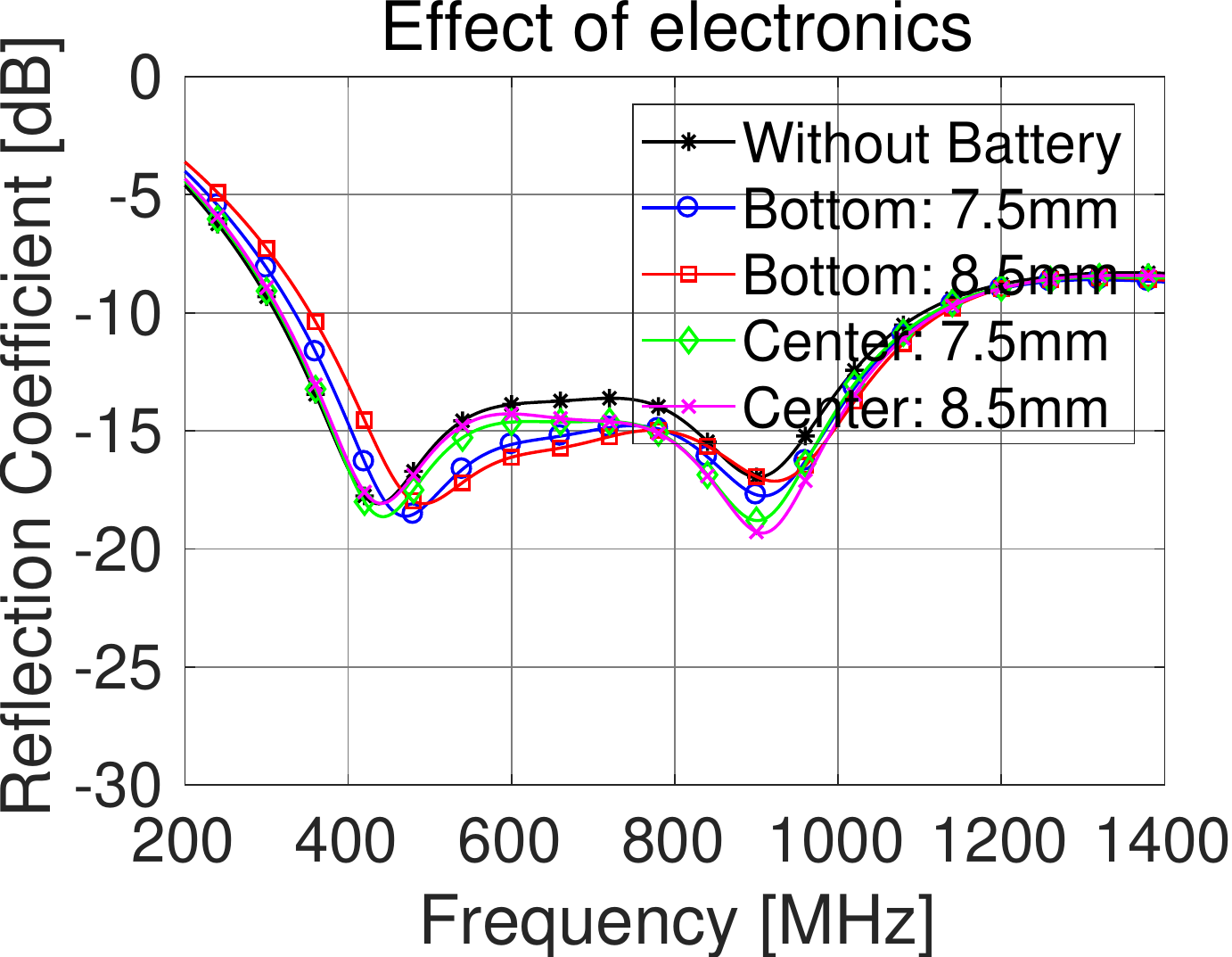}
				\label{fig:battery_effect}}
			\subfigure[]{\includegraphics[scale=.3]{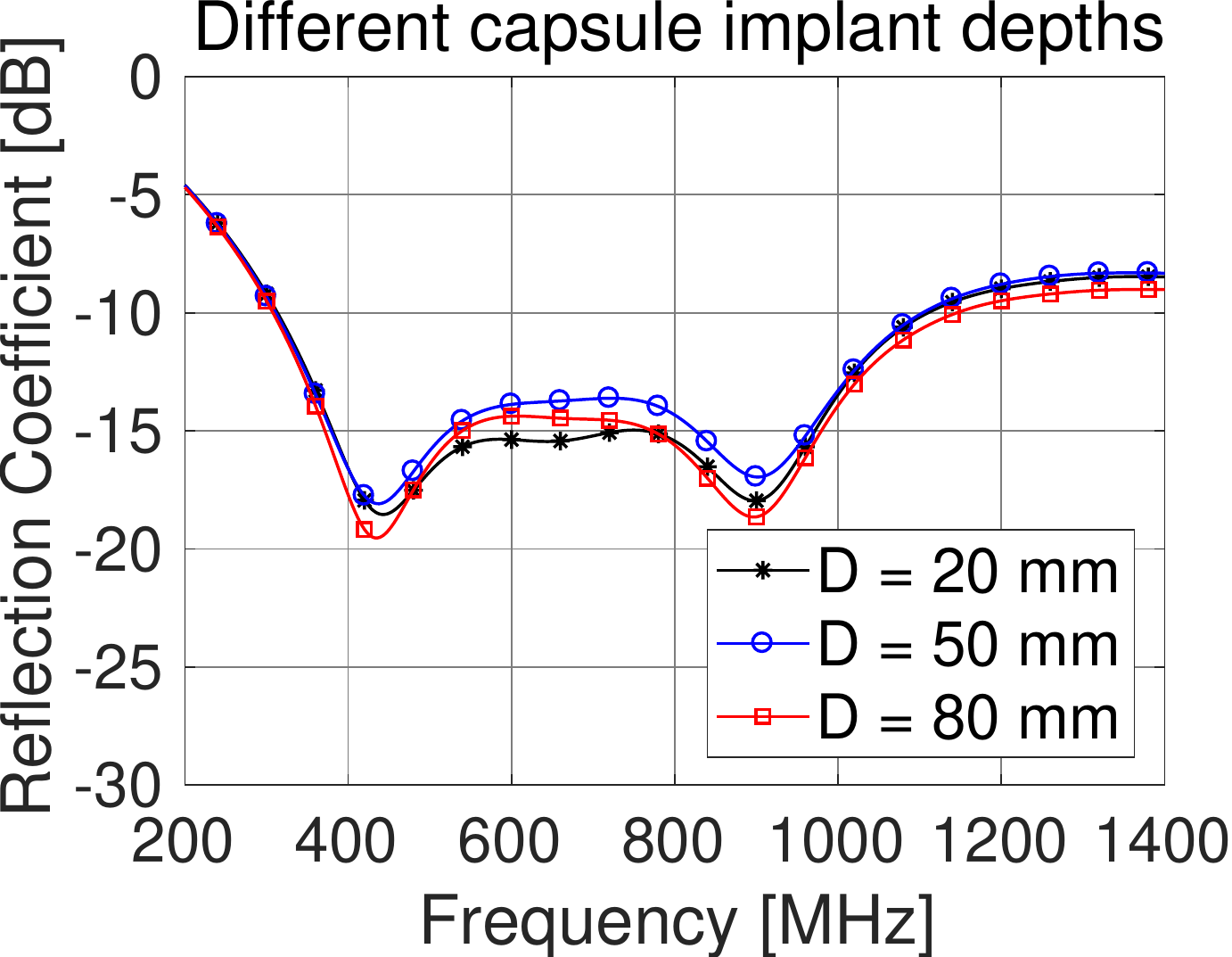}
				\label{fig:depth_z}}
			\subfigure[]{\includegraphics[scale=.3]{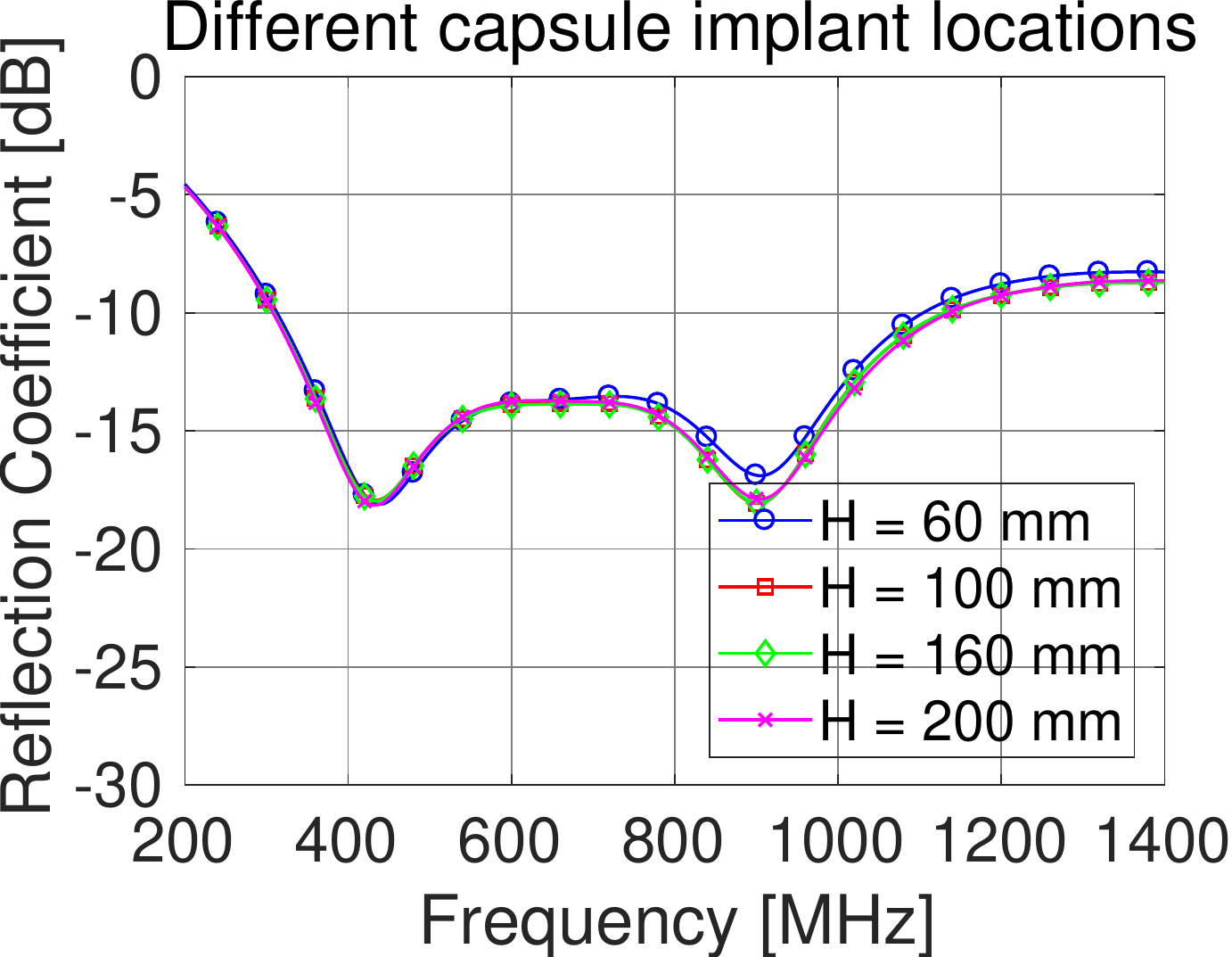}
				\label{fig:depth_side}}
			\subfigure[]{\includegraphics[scale=.3]{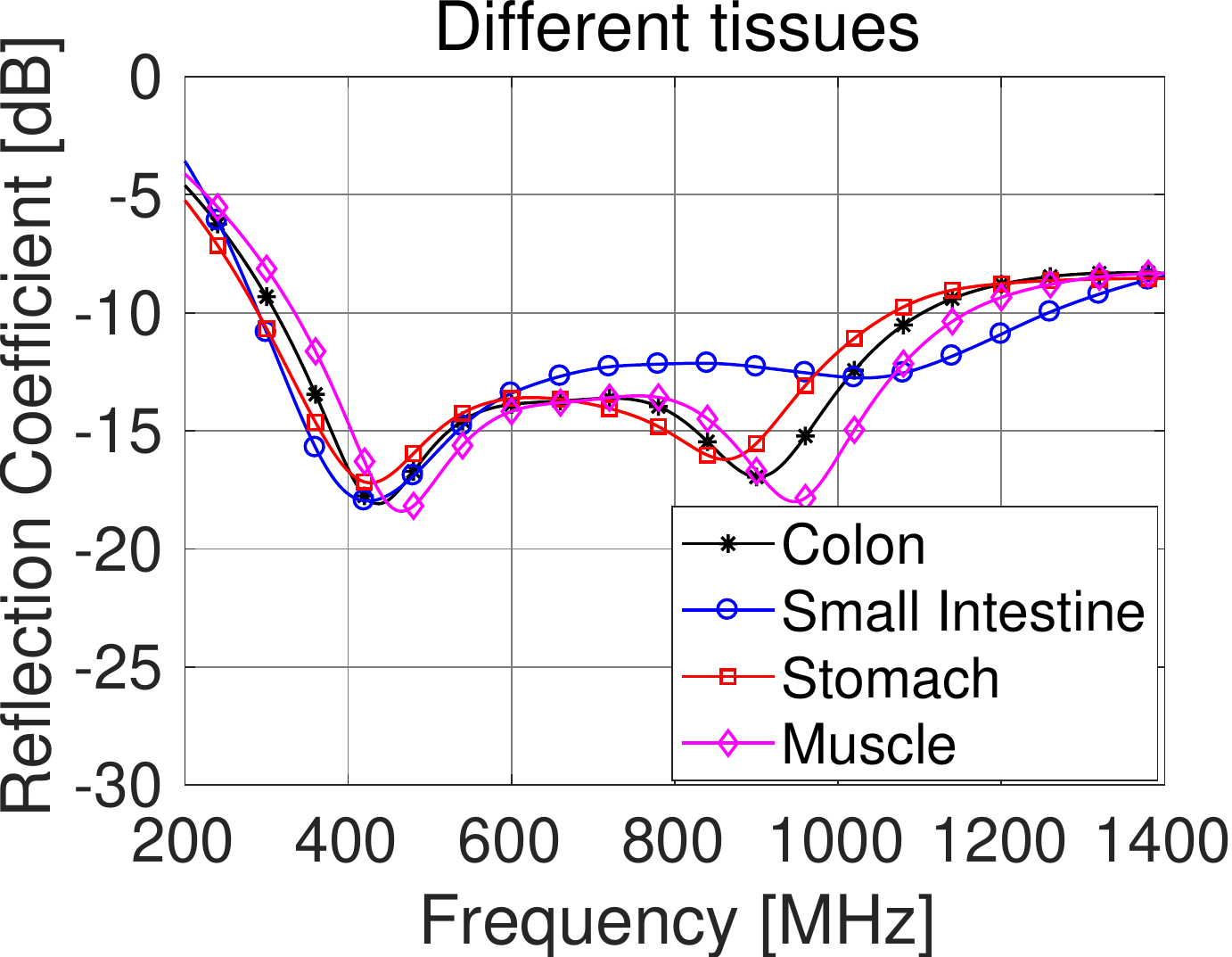}
				\label{fig:different_tissues}}
			\caption{Detuning effects of capsule antenna matching at the center of the colon-tissue phantom as in Fig. \ref{fig:sim_arrangement} due to (a) capsule orientations, (b) biocompatible material, (c) batteries inside the capsule, (d) capsule depths, (e) capsule locations, and (f) different surrounding tissues.}
			\vspace{-20pt}
			\label{fig:sim_orientation_bio_leyer}
		\end{center}
	\end{figure}

Biocompatibility is the property of materials that does not cause any toxic reactions, effects, or injuries in the human body. Because the antennas are made from non-biocompatible materials, an introduction of a biocompatible insulation on the antenna is inevitable in practice. The materials also isolate the capsule from the moist and corrosive environment inside the GI tract. A popular biocompatible material for capsules is a thin layer of low-loss Polyamide with $\varepsilon_{\rm r}=4.3$ and $\tan\delta =0.004$ \cite{Merli_thesis}. A $Y$-oriented capsule antenna with $0.1$~mm thick Polyamide layer was simulated at the center of the colon-tissue phantom shown in Fig.~\ref{fig:sim_arrangement}.  Fig.~\ref{fig:sim_with_and_with_bio_layer} shows the $|S_{11}|$ of the capsule antenna with and without biocompatible layer. Results indicate that the antenna resonance shifts slightly up in frequency, because of lower dielectric loading effect of the antenna due to the biocompatible layer. However, $|S_{11}|$ remains below $-10$~dB between $400$ and $600$~MHz and maintains UWB characteristic. 
\vspace{-10pt}	

\subsubsection{Effects of Electronic Components in the Capsule}
\label{sec:effect_of_electronics}
The capsule transmitter comprises of electronic components, such as illuminating light, telemetry unit, camera, and battery, among which the battery occupies the largest volume. The effect of these electrical components on $|S_{11}|$ was numerically simulated for varying sizes of the battery and its position inside the capsule. The battery was modeled as a cylinder made of a perfect electric conductor with $7.5$ or $8.5$~mm diameters and $7.2$~mm height and was placed at either the bottom or center of the capsule. Results are shown in Fig.~\ref{fig:battery_effect} indicating that the battery with $8.5$~mm diameter slightly increases the resonance frequency of the antenna compared to a hollow capsule. The antenna still maintained $|S_{11}|$ lower than $-10$ dB and a wide-enough matching bandwidth around $433$ MHz.	
\subsubsection{Effects of Capsule Locations and Depths}
\label{sec:imp_loc_dep}
We studied the detuning of antenna's resonance frequency at different locations and depths within the phantom. The antenna implant depth, $D$, was first changed along \textit{Z}-axis from top to bottom, while maintaining $H$ = $117.5$~mm from the left wall of the phantom, as shown in Fig.~\ref{fig:sim_arrangement}. Similarly, the antenna location, $H$, was changed along \textit{Y}-axis, while maintaining $D$ = $50$ mm. The simulated $|S_{11}|$ are presented in Fig.~\ref{fig:depth_z} and Fig.~\ref{fig:depth_side}, showing that they are practically unchanged for all the tested depths and locations.      	
\subsubsection{Effects of Different Tissues}
\label{sec:effect_of_different_tissue}

As a capsule travels through the entire GI tract, it experiences a significant change of relative permittivity and conductivity depending on the surrounding tissues, as presented in Table~\ref{table:different_tissue} \cite{C_Gabriel}. Fig.~\ref{fig:different_tissues} shows $|S_{11}|$ for different surrounding tissues of the phantom in Fig.~\ref{fig:sim_arrangement}, indicating slight decrease of the antenna's resonance frequency in the stomach and small intestine due to the higher $\varepsilon_{\rm r}$ than the colon. Furthermore, matching bandwidth is enhanced when the antenna is in the small intestine. Since, the quality factor of the capsule antenna in the small intestine is the smallest due to the highest $\tan \delta$, the antenna shows the largest matching bandwidth among other tissues. As the muscle tissue phantom is frequently used for the performance evaluation of capsule antennas in literature \cite{E_FMerliTAP,c_denys_TAP2017,c_JFaerber_Tcir_2017,Kiourti_P,Izdebski_TAP,c_rajagopalan}, we evaluated the antenna matching in muscle tissue as well. The antenna in muscle tissue resonates at a slightly higher frequency due to lower $\varepsilon_{\rm r}$. Despite clear variations in the antenna resonance across tissue types, $|S_{11}|$ remains below $-10$~dB between $336$ and $1065$ MHz in all cases.
	\vspace{-15pt}  
%Surrounded by different tissues with varying dielectric parameters as summarized in Table~\ref{table:different_tissue}, the capsule antenna detunes during endoscope operation. 
	\begin{table}
		\begin{center}
			{\caption{The dielectric and conductive properties of stomach, colon and small intestine at $433$ MHz \cite{C_Gabriel}} 
				\label{table:different_tissue}}
			\begin{tabular}{c|c|c|c} \hline
				Tissues  & $\varepsilon_{\rm r}$  & $\sigma$ [S/m]& $\tan \delta$\\
				
				\hline
				
				Stomach & 67.2 & 1.01  & 0.62\\
				Colon & 62.0 & 0.87  & 0.58\\
				Small Intestine & 65.2 & 1.92 & 1.22 \\
				Muscle & 56.9 & 0.8  & 0.59 \\
				\hline
			\end{tabular}
		\end{center}
		\vspace{-10pt}
	\end{table}
	%In the previous study, a single-layer colon phantom was utilized to design and optimize the capsule antenna. A similar single-layer colon phantom was used for the prototype measurements, so the measurement results will be compared with the according simulated results to validate the antenna design. However, it would be more accurate if the proposed antenna was also evaluated in a realistic human body model. In order to study
\subsection{Resonance, Radiation, Specific Absorption Rate (SAR) in a Realistic Human Body Model}
\label{sec:capsule_in_human_body_phantom}
\begin{figure}[t!]
	\begin{center}
		\resizebox{0.80\hsize}{!}{\includegraphics[scale=.45]{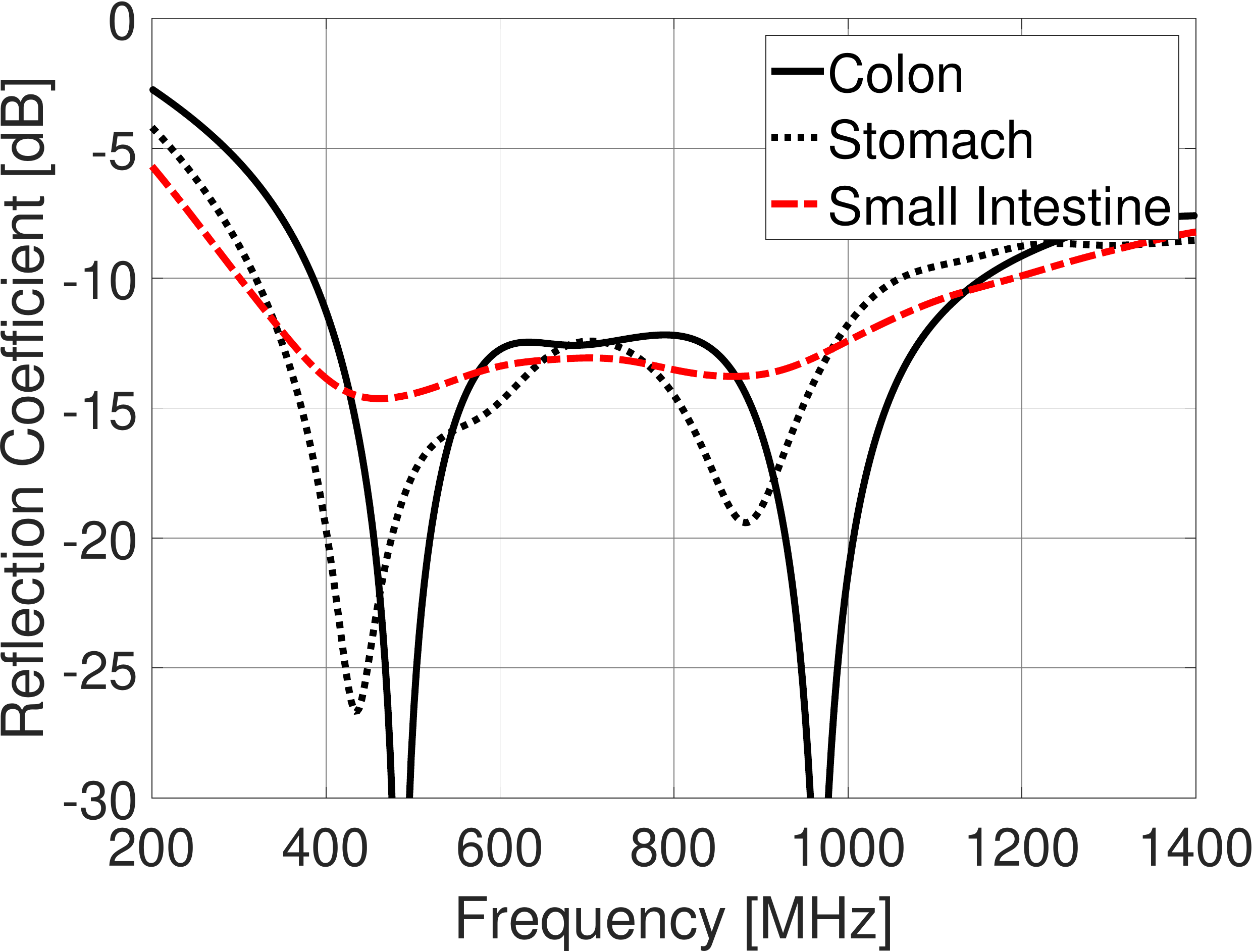}}
\caption{ Simulated reflection coefficient of the capsule antenna in CST Gustav voxel human body with three different positions, i.e., in colon, stomach and small intestine.}
		\vspace{-20pt}
		\label{fig:sim_s11_anatomical_tissue}
	\end{center}
\end{figure}

The resonance and radiation performance of the capsule antenna was studied with 3-D CST Gustav voxel human body model. Due to the limited computing resources, only a torso with the volume $290  \times  230  \times  100$ $\rm mm^3$ was considered, which is comparable with the dimensions of the single-layer colon phantom shown in Fig.~\ref{fig:sim_arrangement}. When implanted in the colon, stomach, and small intestine, the capsule is $50$, $90$ and $85$ mm away from the nearest body surface, respectively. The simulated $|S_{11}|$ of the optimized capsule antenna without electronics and biocompatible layer with three implant positions are presented in Fig.~\ref{fig:sim_s11_anatomical_tissue}. The results demonstrate that for the proposed antenna $|S_{11}|$ is better than $-10$ dB at $433$ MHz for all implant positions and maintains a wide impedance matching bandwidth. A comparison of the results in Fig.~\ref{fig:sim_s11_anatomical_tissue} with those in Fig.~\ref{fig:different_tissues} indicates the suitability of using the simplified colon phantom in Fig.~\ref{fig:sim_arrangement} for the capsule antenna design. The simulated peak realized gain is $-23.5$, $-26.8$, and $-29.7$ dBi for the capsule antenna implanted in the colon, stomach, and small intestine, respectively. The antenna in small intestine shows a lower peak gain due to the higher conductivity compared to colon and stomach. Here, peak gain is also affected by the distance from body surface.

%\textcolor{red}{Fig.~\ref{fig:gain_ana_colon_Stomach_SI} shows of the simulated 3-D radiation pattern of the proposed capsule antenna when antenna is implanted in three different positions as shown Fig.~\ref{fig:TAP_1_capsule_anatomicalai}. Results depict that the realized peak gain is $-23.5$ dBi, $-26.8$ dBi and $-29.7$ dBi when capsule antenna implanted in colon, stomach and small intestine, respectively. Antenna in small intestine shows lower peak gain due to the higher conductivity compared to colon and stomach. However, in general peak gain is affected by the properties of the surrounding tissue and the distance to the body surface.}  
%	\begin{figure}[t!]
%	\begin{center}
%		\subfigure[]{\includegraphics[scale=.35]{fig/TAP_1_gain_colon_ana.eps}
%	\label{fig:TAP_1_gain_colon_ana}}
%		\subfigure[]{\includegraphics[scale=.35]{fig/TAP_1_gain_Stomach_ana.eps}
%			\label{fig:TAP_1_gain_stomach_ana}}
%		\subfigure[]{\includegraphics[scale=.35]{fig/TAP_1_gain_SI_ana.eps}
%			\label{fig:TAP_1_gain_SI_ana}}
%		\caption{\textcolor{red}{Simulated 3-D radiation pattern (realized gain) using CST at $433$ MHz when proposed antenna is implanted in three different positions in Fig.~\ref{fig:TAP_1_capsule_anatomicalai}. Antenna implanted in the (a) Colon and (b) Stomach and (c) Small intestine.}}
%		\vspace{-10pt}
%		\label{fig:gain_ana_colon_Stomach_SI}
%	\end{center}
%\end{figure}

\label{sec:SAR_inbody}
Since the capsule antenna needs to be swallowed in a human body, radiation safety should be discussed. There is a maximum allowable power radiated from the capsule antenna. The SAR is the rate of energy deposited per unit mass of tissue. The IEEE C95.1-1999~\cite{SAR_1999} and C95.1-2005 \cite{SAR_2005} specify that 1-g and 10-g averaged SAR should be less than 1.6 W/kg and 2 W/kg, respectively. The SAR calculator in CST Microwave Studio therefore numerically estimated the maximum allowable input powers to the $Y$-orientated capsule antenna implanted at three different positions in the CST Gustav voxel human body model. The results demonstrate that the capsule antenna is safe to be used at the transmit power less than $7.1$ and $28$~mW in colon, $5.0$ and $24$~mW in small intestine and $7.2$ and $25$~mW in stomach for the 1-g and 10-g averaged SAR, respectively. The capsule for wireless endoscopy in mind is likely to fulfill the SAR requirement.

	\vspace{10pt}

%	\begin{table}
%	\begin{center}
%		{\caption{Simulated Maximum SAR (input power = 1 W), and maximum input power for satisfying the SAR requirement in the human body at $433$ MHz} 
%			\label{table:SAR_table}}
%	%	\begin{tabular}{c|c|c|c|c} \hline
%		\begin{tabular}{m{1.0cm}|m{1.4cm}|m{1.45cm}|m{1.4cm}|m{1.45cm} } \hline
					%	\begin{tabular}{p{0.05\linewidth}p{0.25\linewidth}p{0.07\linewidth}p{0.08\linewidth}p{0.1\linewidth}p{0.28\linewidth}}
%			Implant position  & max. 1-g avg. SAR [W/Kg]  & max. input power [mW]& max. 10-g avg. SAR [W/Kg]& max. input power [mW]\\
			
%			\hline
%			Colon & 223 & 7.1  & 72.3& 28\\
%				\hline
						
%			Small Intestine & 332 & 5.0 & 81.7& 24 \\
%				\hline
%			Stomach & 222 & 7.2  & 79.1& 25\\
%				\hline

%		\end{tabular}
%	\end{center}
%	\vspace{-10pt}
%\end{table}

\section{On-body Antenna Designs and Simulations}
\subsection{Antenna Structure}
\label{sec:on_body_antenna_structure}
\begin{figure}[t!]
		\begin{center}
			\subfigure[]{\includegraphics[scale=.52]{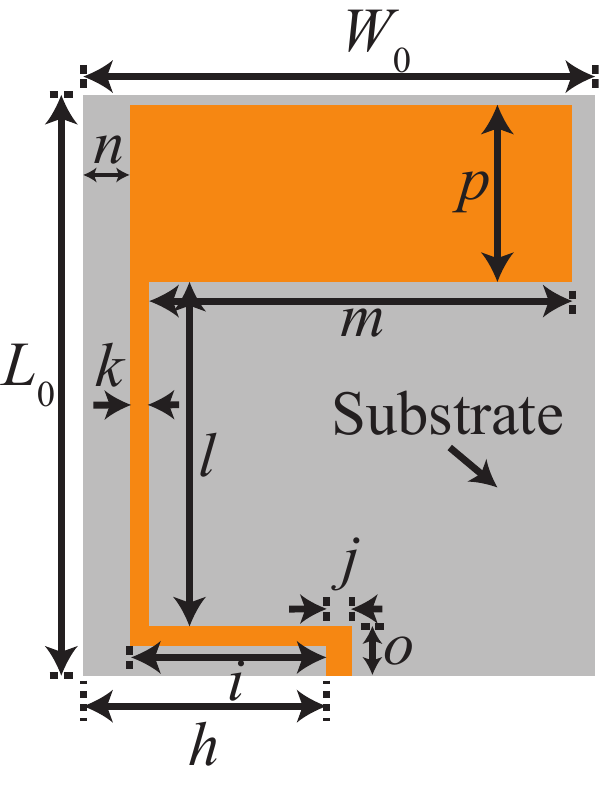}
				\label{fig:on_body_top_view}}
			\subfigure[]{\includegraphics[scale=.55]{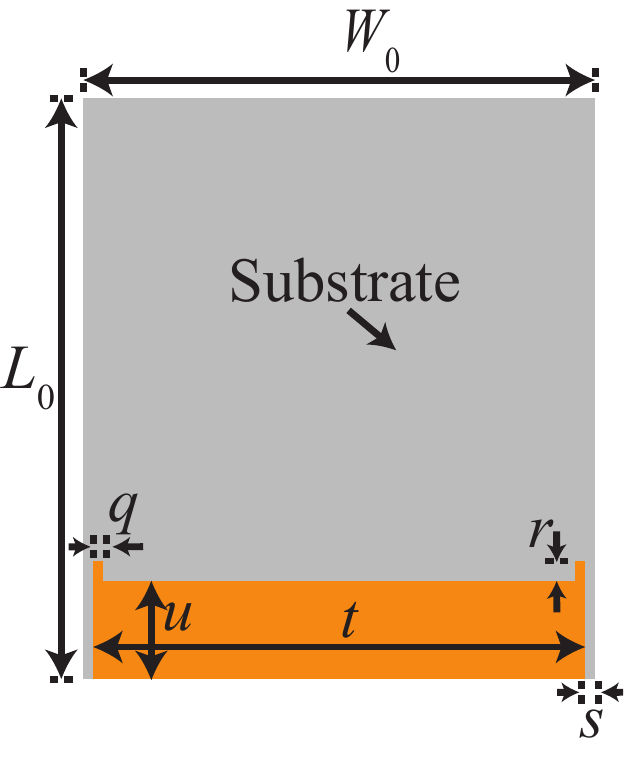}
				\label{fig:on_body_back_view}}
			\caption{On-body antenna structure: (a) top view, and (b) bottom view.}
			\vspace{-10pt}
			\label{fig:on_body_antenna_structure}
		\end{center}
	\end{figure}

The design goal of the on-body antenna is to have a compact size, and sufficient resonance and radiation characteristics at $433$~MHz. Our on-body antenna solution consists of a compact meandered monopole antenna, along with two small wings on a ground plane and partial grounding, which improve antenna matching. The configuration of the on-body antennas is shown in Fig.~\ref{fig:on_body_antenna_structure}. A $1.5$~mm thick FR4 with $\varepsilon_{\rm r} = 4.3$ is used as substrate material, whereas $19~ \rm{\mu m}$ thick copper is used as conductor material to pattern the monopole and ground plane. The optimized dimensions of the antenna are summarized in Table~\ref{table:on_body_antenna_dimensions}.
\vspace{-15pt}       
	
	\begin{table}[t!]
		\begin{center}
			{\caption{Optimized dimensions of the on-body antenna structure in mm} 
				\label{table:on_body_antenna_dimensions}}
			\begin{tabular}{m{.2cm}m{.2cm}m{.25cm}m{.12cm}m{.12cm}m{.05cm}m{.05cm}m{.12cm}m{.12cm}m{.05cm}m{.12cm}m{.05cm}m{.05cm}m{.05cm}m{.12cm}m{.1cm}} \hline
				%\begin{tabular}{l*{14}{c}r}
				$L_o$  & $W_0$  & $h$ & $i$ & $j$ & $k$ & $l$ & $m$ & $n$ & $o$ & $p$ & $q$& $r$ & $s$ & $t$& $u$\\
				
				\hline
				59 & 52 & 24.7 & 20 & 2.6 & 2 & 35 & 43 & 4.7& 5 & 18 & 1 & 2 & 1& 49 & 10 \\
				
				\hline
			\end{tabular}
		\end{center}
		\vspace{-15pt}
	\end{table} 
	\subsection{Effects of Body on Matching and Radiation}
	In practice, antennas of the receiver unit of a WCE system are placed directly on the human body. We studied the influence of the tissue on resonance and radiation characteristics of the proposed on-body antenna using the single-layer colon tissue phantom introduced in Section~\ref{sec:capsule_implementation_and_operating_environments} and the 3-D CST Gustav voxel human body used in Section~\ref{sec:capsule_in_human_body_phantom}. The simulation results of the antenna on a single-layer phantom will be compared to corresponding phantom measurements in Section~\ref{sec:measurements_on_body_antenna}, while results with the antenna on anatomical human body model will be validated through \textit{ex-vivo} measurements on a test person.
	
	\begin{figure}[t!]
		\begin{center}
			\resizebox{0.9\hsize}{!}{\includegraphics[scale=.8]{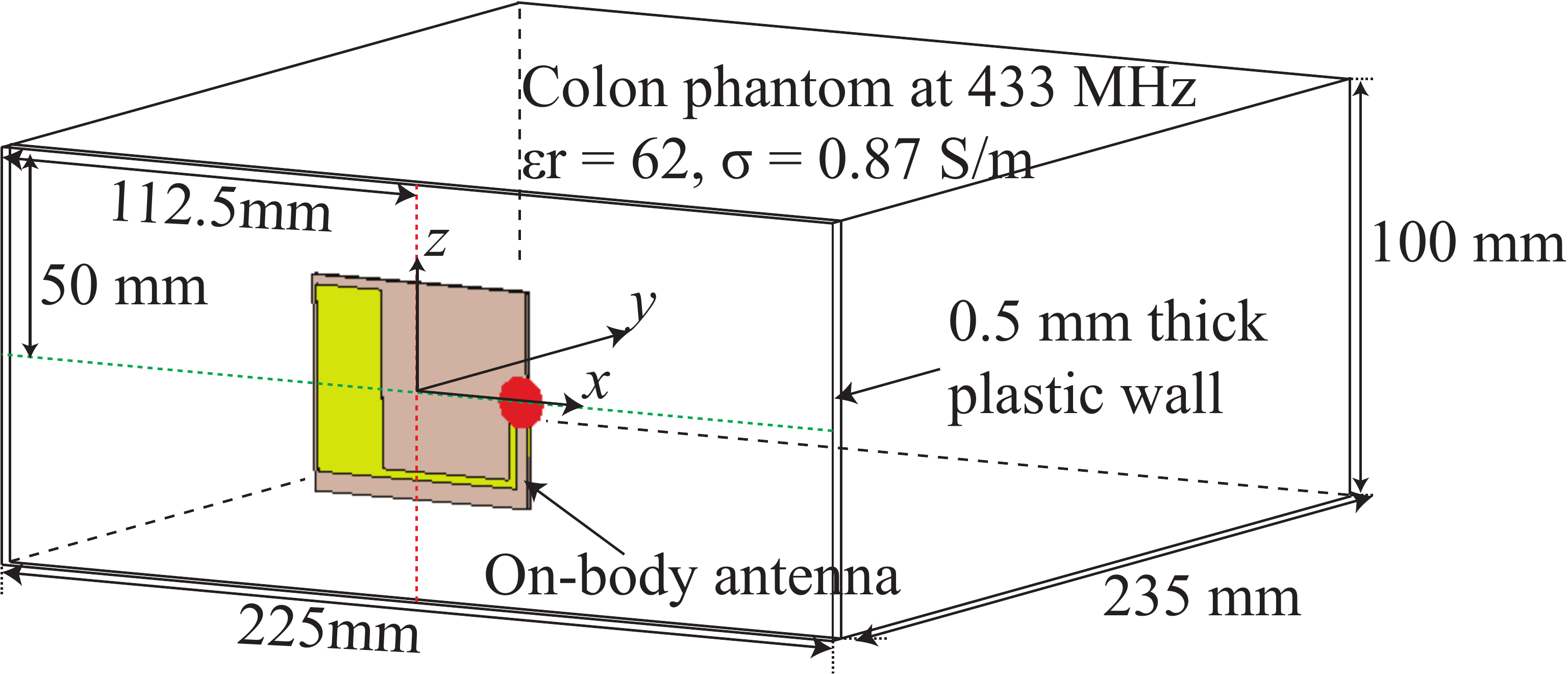}}
			\caption{Receiver antenna on a single-layer colon tissue phantom.}
			\vspace{-10pt}
			\label{fig:sim_arrangement_on_body}
		\end{center}
	\end{figure}
	
	\subsubsection{Single Tissue Phantom}
	\label{sec:on_body_antenna_single_tissue_phantom}
\begin{figure}[t!]
		\begin{center}
			\resizebox{0.80\hsize}{!}{\includegraphics[scale=.45]{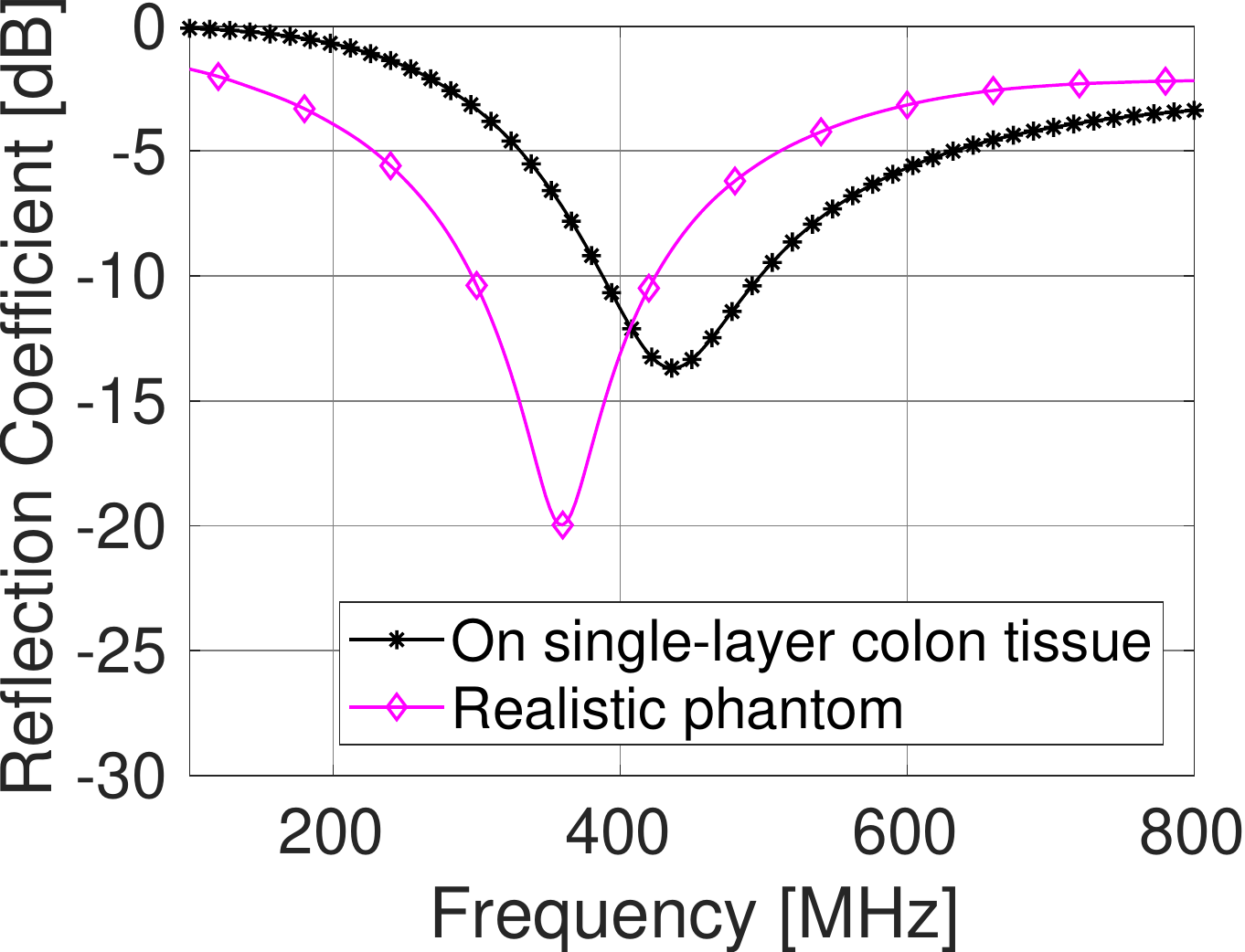}}
			\caption{Simulated reflection coefficient of the on-body antenna.}
			\vspace{-20pt}
			\label{fig:S11_on_body_simulation}
		\end{center}
	\end{figure}

	The antenna was placed in the center of the outer wall of the colon-tissue
	phantom model as shown in Fig.~\ref{fig:sim_arrangement_on_body}. For consistency with set-ups of the measurement described in Section IV-A2, a $0.5$~mm thick plastic container with $\varepsilon_{\rm r}=1.88$ and $\tan\delta =0.005$ was included in the simulation model. An SMA edge connector was used as the antenna feed, but it prevented the antenna from being fully touching the surface of the phantom. So, this gap of $1$~mm between the antenna and plastic cover was introduced also in the simulations. The simulated $|S_{11}|$ shown in~Fig.~\ref{fig:S11_on_body_simulation} illustrates that the antenna resonates at $435$ MHz with $-10$~dB impedance matching across $110$ MHz bandwidth. The peak realized gain is $-18$ dBi. 
	
\subsubsection{Realistic Human Body Model}
\label{sec:on_body_antenna_multi_tissue_phantom}
In addition, the resonance and radiation performance of the antenna was evaluated using a 3-D CST Gustav voxel human body model. The dimensions of the human body torso used in the numerical simulations are same as in Section~II-C. According to \cite{Position_orientation_Chirwa}, the antenna was placed directly on the left-side abdominal skin of the torso. The simulated $|S_{11}|$ shown in~Fig.~\ref{fig:S11_on_body_simulation} depicts that the antenna resonates at $360$ MHz. The downward shift of the resonance frequency is reasonable since the antenna was in direct contact with the skin tissue. The peak realized gain of the antenna is $-16.5$ dBi. Finally, the SAR was estimated using CST Microwave Studio. We found that the maximum allowable input powers to the  antenna placed on the abdomen of CST Gustav human body model is $18$ and $82$~mW for the 1-g and 10-g averaged SAR, respectively.

%	\begin{figure}[t!]
%	\begin{center}
%		\resizebox{0.80\hsize}{!}{\includegraphics[scale=.45]{fig/TAP_1_On_body_anatomical_sim_env.eps}}
%		\caption{On-body antenna position on the abdomen of a 3-D CST Gustav voxel human body model.}
%		\vspace{-10pt}
%		\label{fig:TAP_1_On_body_anatomical_sim_env}
%	\end{center}
%\end{figure}

\section{Antenna Fabrication and Measurements}
\subsection{Capsule Antenna}
\subsubsection{Fabrication}
\begin{figure}[t]
		\begin{center}
			\subfigure[]{\includegraphics[scale=.4]{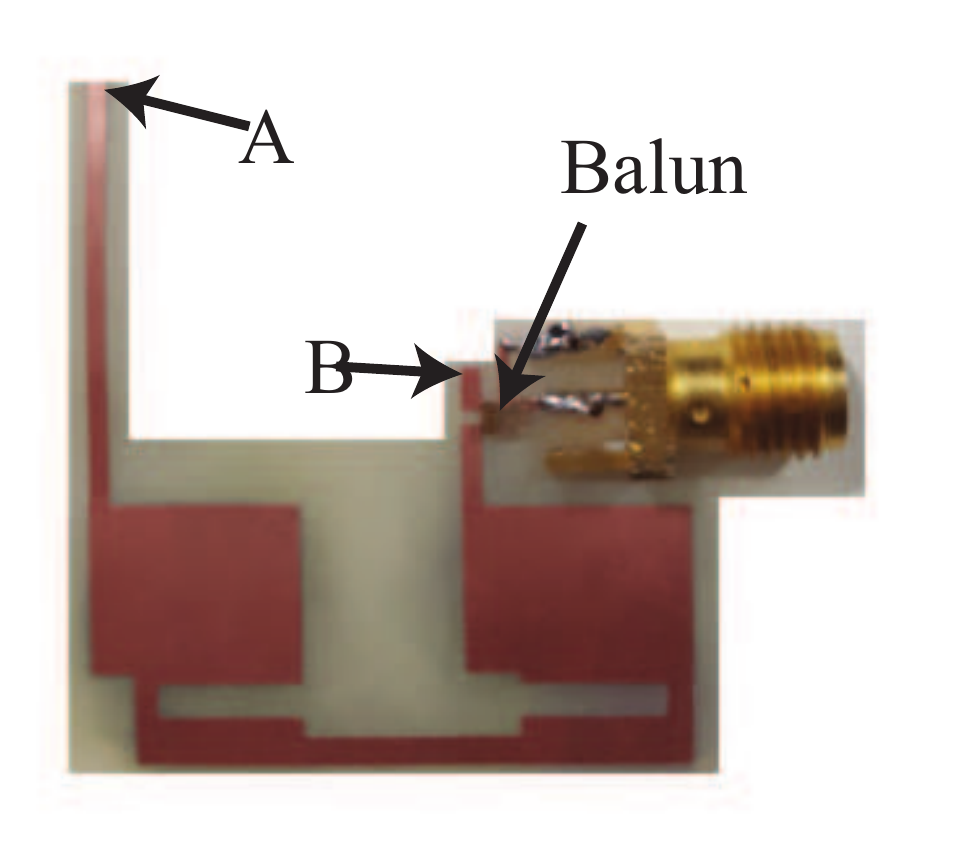}
				\label{fig:protoype1}}
			\subfigure[]{\includegraphics[scale=.2]{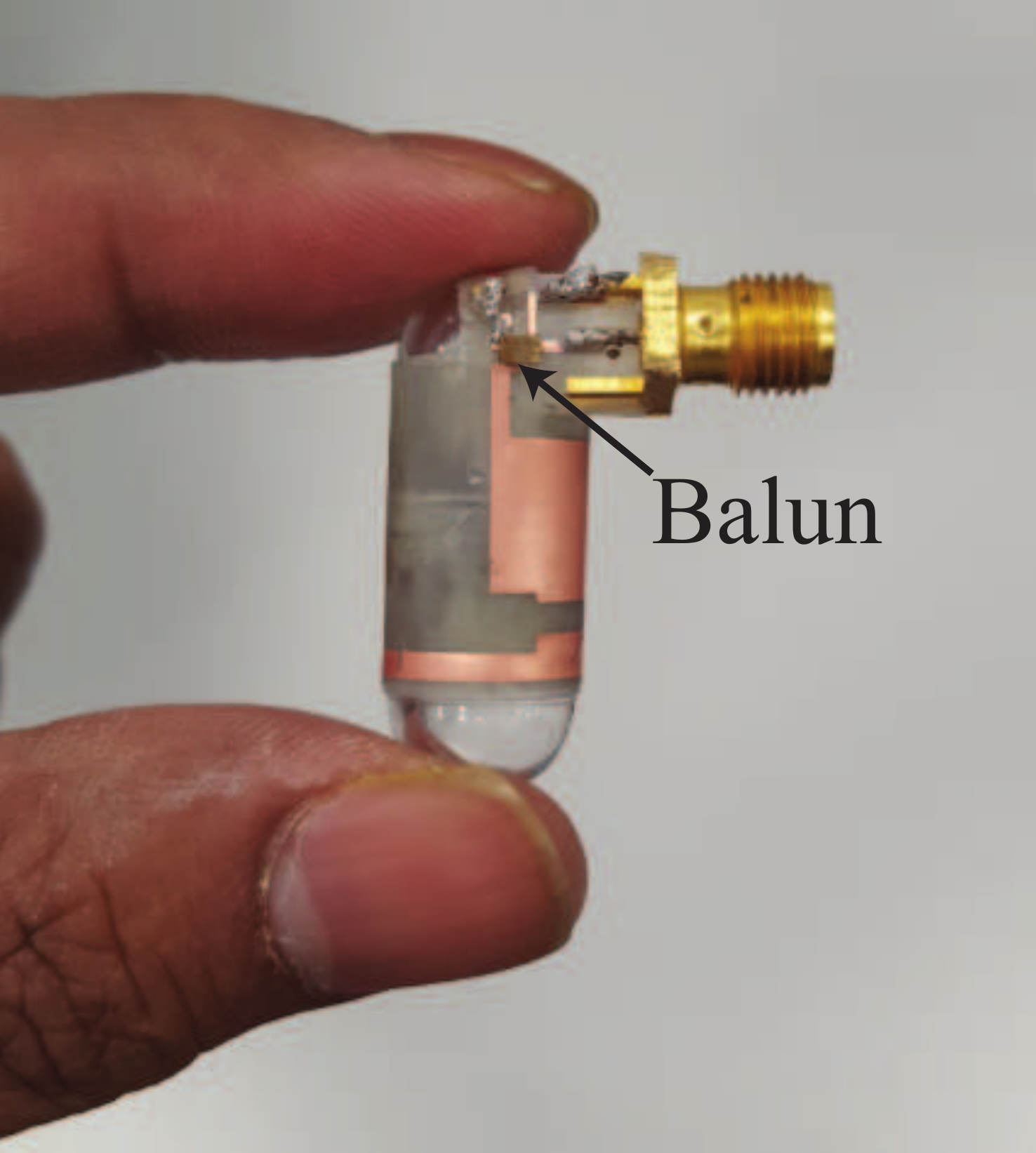}
				\label{fig:in_body_antenna_warp_around_capsule}}
%			\subfigure[]{\includegraphics[scale=.55]{fig/capsule_module.eps}
%				\label{fig:capsule_module}}
			\caption{Fabricated capsule antenna (a) before, and (b) after placed around the capsule.}
			\vspace{-20pt}
			\label{fig:fabricated_antenna_prototype}
		\end{center}
	\end{figure}

The fabricated capsule antenna before wrapping on the capsule module is presented in Fig.~\ref{fig:protoype1}. Since a broadband surface mount balun and an SMA connector will be used to feed the antenna in the \textit{in-vitro} measurement, the footprints of the balun and the SMA connector were also fabricated along with the antenna. The antenna was fabricated on the same flexible substrate material as defined in Section~\ref{sec:capsule_antenna_structure}; details of the fabrication process are described in~\cite{Suzan_eucap}. The dimensions of the flat capsule antenna follow in Table~\ref{table:antenna_dimensions}, and it was wrapped around the outer-wall of a standard capsule module as illustrated in Fig.~\ref{fig:in_body_antenna_warp_around_capsule}. The capsule module is made of polystyrene with $\varepsilon_{\rm r} = 2.6$ and $\tan \delta = 0.05$ at $1$ GHz. The diameter and length of the capsule are $11$~mm and $27$ mm, respectively, whereas the thickness of the wall is $0.5$~mm. After wrapping the points A and B in Fig.~\ref{fig:protoype1} were soldered together to form the loop. The biocompatible layer and any components inside the capsule were not included in the measurement. 
	
	\subsubsection{Measurement Set-up}
	\label{sec:meas_phantom}
\begin{figure}[t!]
		\begin{center}
			\resizebox{0.99\hsize}{!}{\includegraphics[scale=.4]{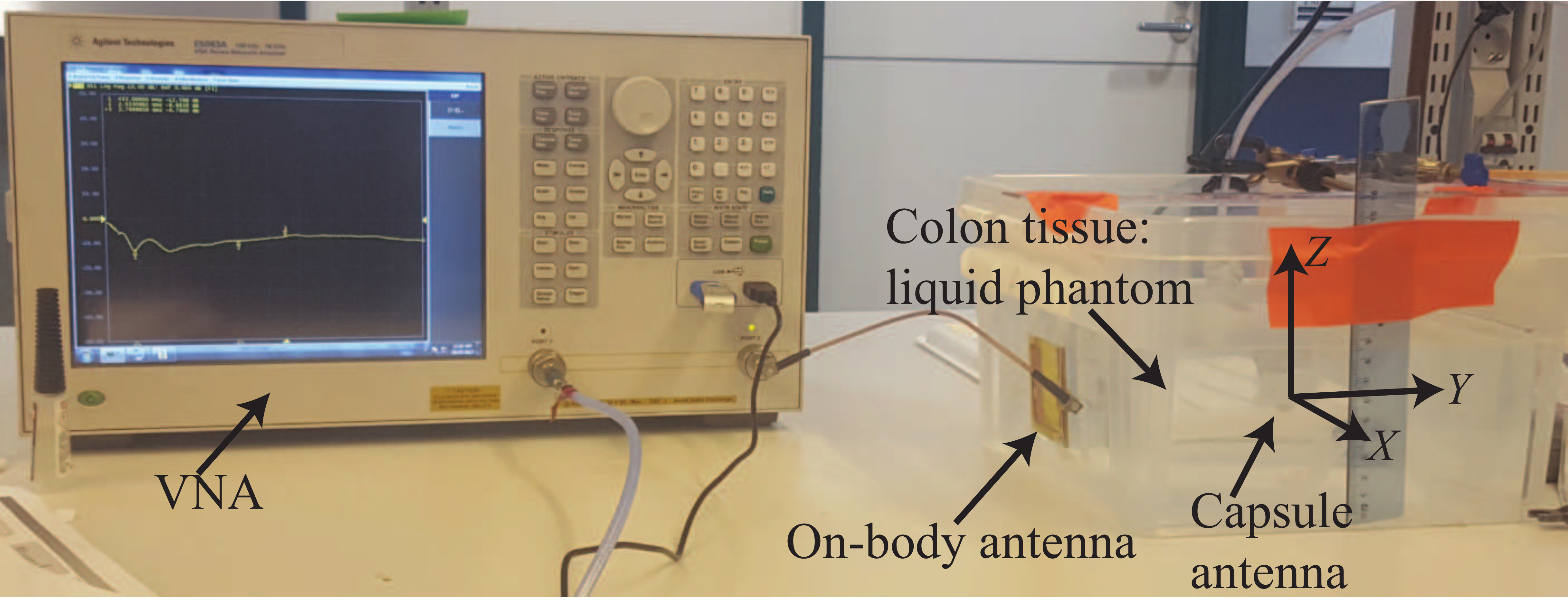}}
			\caption{Illustration of the matching and path loss measurement setups.}
			\vspace{-10pt}
			\label{fig:measurement_setup}
		\end{center}
	\end{figure}

The \textit{in-vitro} measurement set-up for the capsule antenna is illustrated in Fig.~\ref{fig:measurement_setup}, which consisted of a vector network analyzer (VNA) and a rectangular-shaped plastic container to form a phantom. The container was filled with liquid mimicking the colon tissue. The liquid was formulated by mixing $79$\% salted water and $21$\% TritonX-100. The mixing process consists of the following steps: dissolving the salt in distilled water as $7.85$~g/L, and heating the TritonX-100 and salted water at $40^\circ$C separately before mixing. We used HP 8720C network analyzer and  85070A dielectric probe kit~\cite{Suzan_eucap} based on the transmission line propagation method to measure the electrical properties of the liquid phantom. The measured permittivity and loss tangent values at $433$ MHz were $61.4$ and $0.60$, respectively, emulating the dielectric properties of the colon tissue properly.
	
Since the proposed loop antenna is balanced, the current leakage would be a possible problem when connecting with an unbalanced transmission line, such as, a coaxial cable. We used balun instead of the setup in \cite{Suzan_eucap} to measure the balanced capsule antenna using an unbalanced connector. For differential feeding, we used similar approach reported in \cite{Sumin_Kim_TAP_L}. A wideband surface mount balun (Analen B0322J5050AHF) was used at the feeding point of the antenna. Balance ports of the balun were connected to the antenna, whereas center and outer conductors of an SMA connector were soldered to the unbalanced port and ground of the balun, respectively. During the measurements, we used a layer of sticky rubber with $\varepsilon_{\rm r} = 2.2, \tan \delta = 5.0 \times 10^{-4}$ around the feeding components, i.e., a balun, an SMA connector, and a coaxial cable to avoid the direct contact with the liquid \cite{Merli_thesis, E_FMerliTAP}. As the VNA was calibrated on the reference plane of the SMA connector, capsule antenna with the SMA connector was simulated to test its impact on matching. Since the SMA connector was isolated from the liquid, we found a negligible impact on matching. The SMA connector inside the sticky rubber works almost as in free space at the designed frequency band. The proposed capsule antenna, which was inserted in the colon liquid phantom, was connected to port-1 of the VNA, while the port-2 in Fig.~\ref{fig:measurement_setup} was open-ended in matching measurements.   
	%The port 2 in Fig.~\ref{fig:measurement_setup} was open-ended in matching measurements of the capsule antenna. The very short feed cable attached to the antenna, shown in Fig.~\ref{fig:in_body_antenna_warp_around_capsule} is found to have negligible effects on matching according to our numerical simulations because a significant amount of power fed to the antenna port is dissipated in the lossy tissue and the current does not return to the ground conductor of the feed cable. We therefore excluded the very short feed cable from the short-open-load-thru calibration. 
	
\subsubsection{Results}
\begin{figure}[t]
		\begin{center}
			\subfigure[]{\includegraphics[scale=.30]{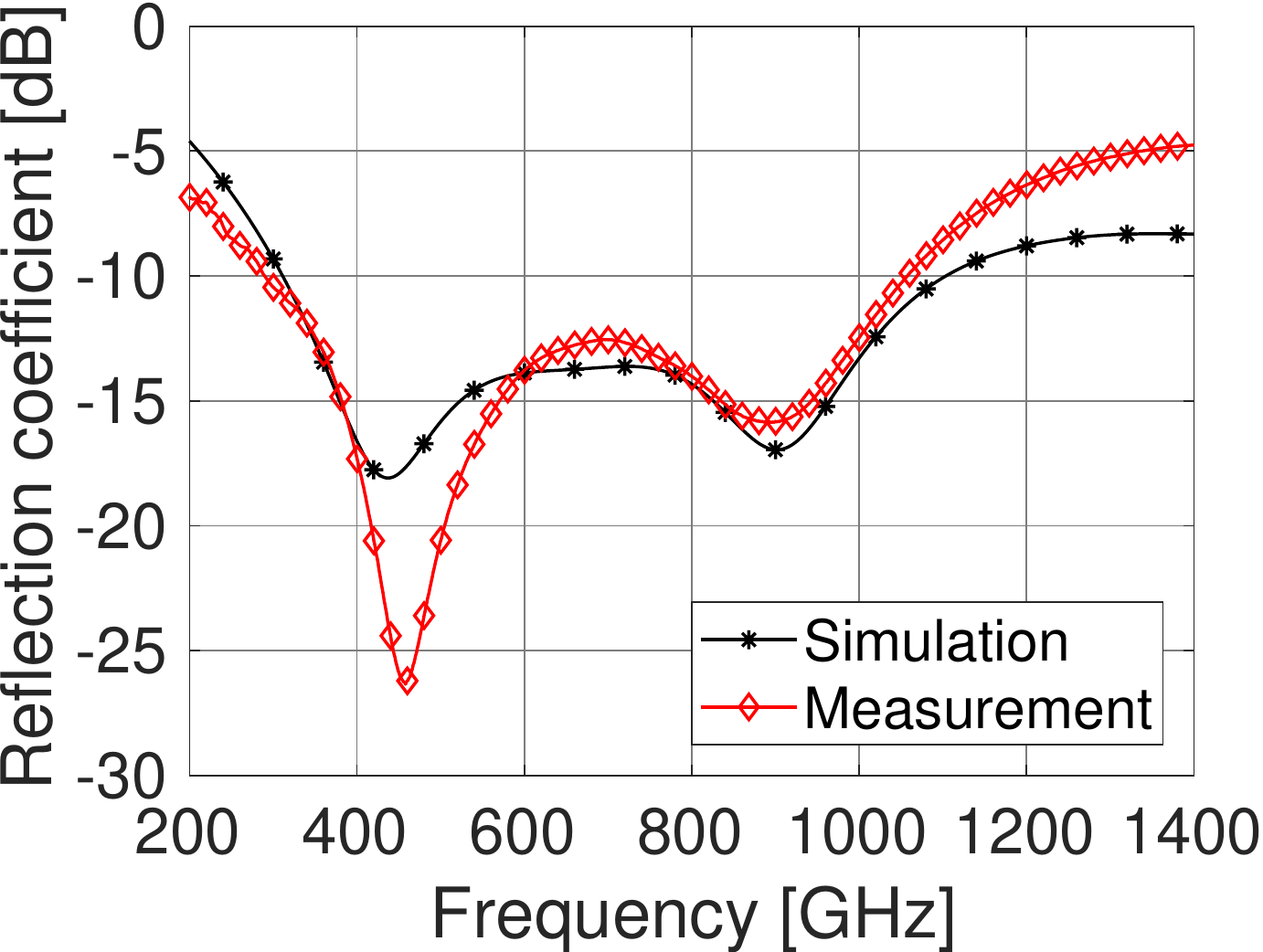}
				\label{fig:sim_measured_antenna_at_center}}
			\subfigure[]{\includegraphics[scale=.225]{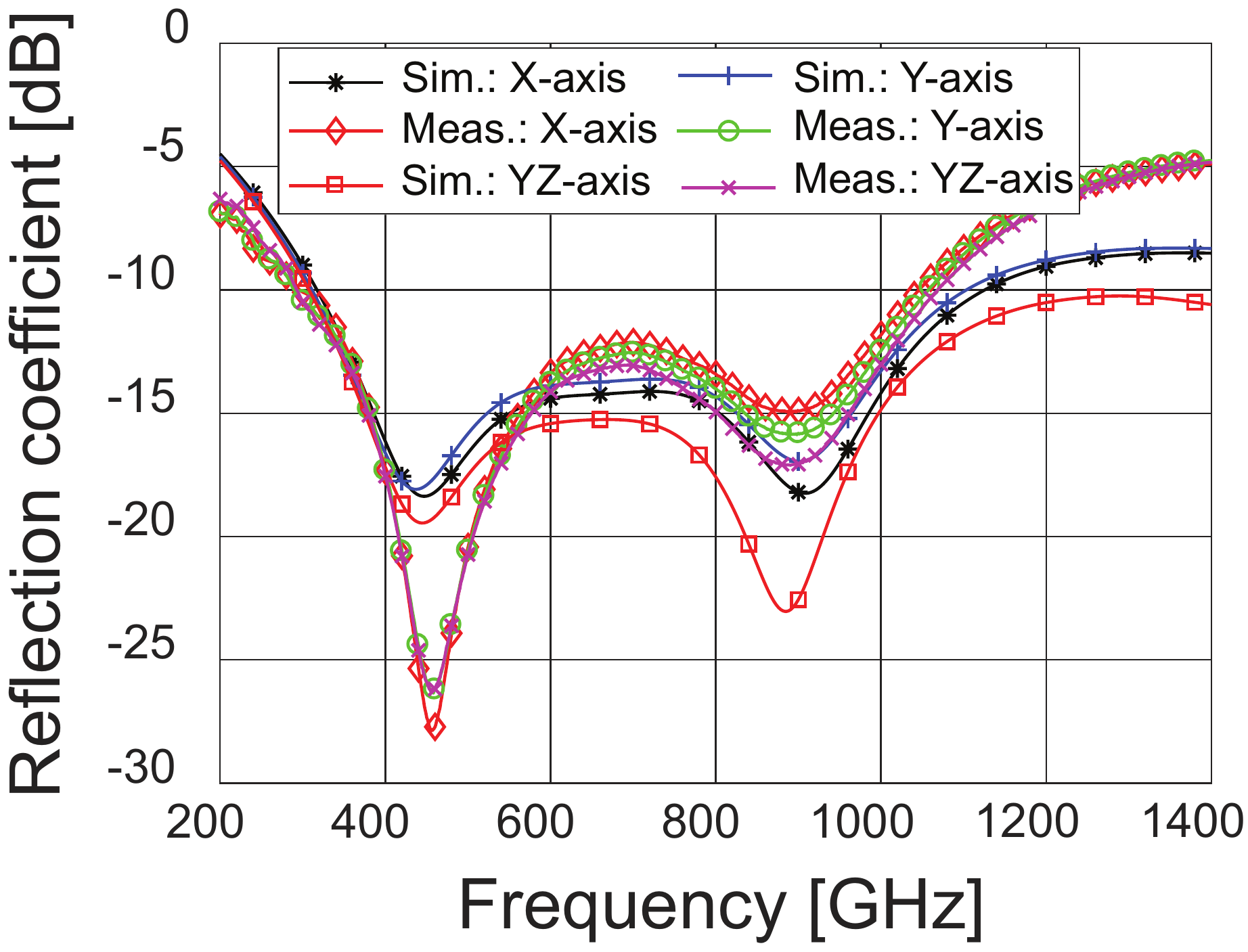}
				\label{fig:meas_in_body_implant_orientations}}
			\subfigure[]{\includegraphics[scale=.30]{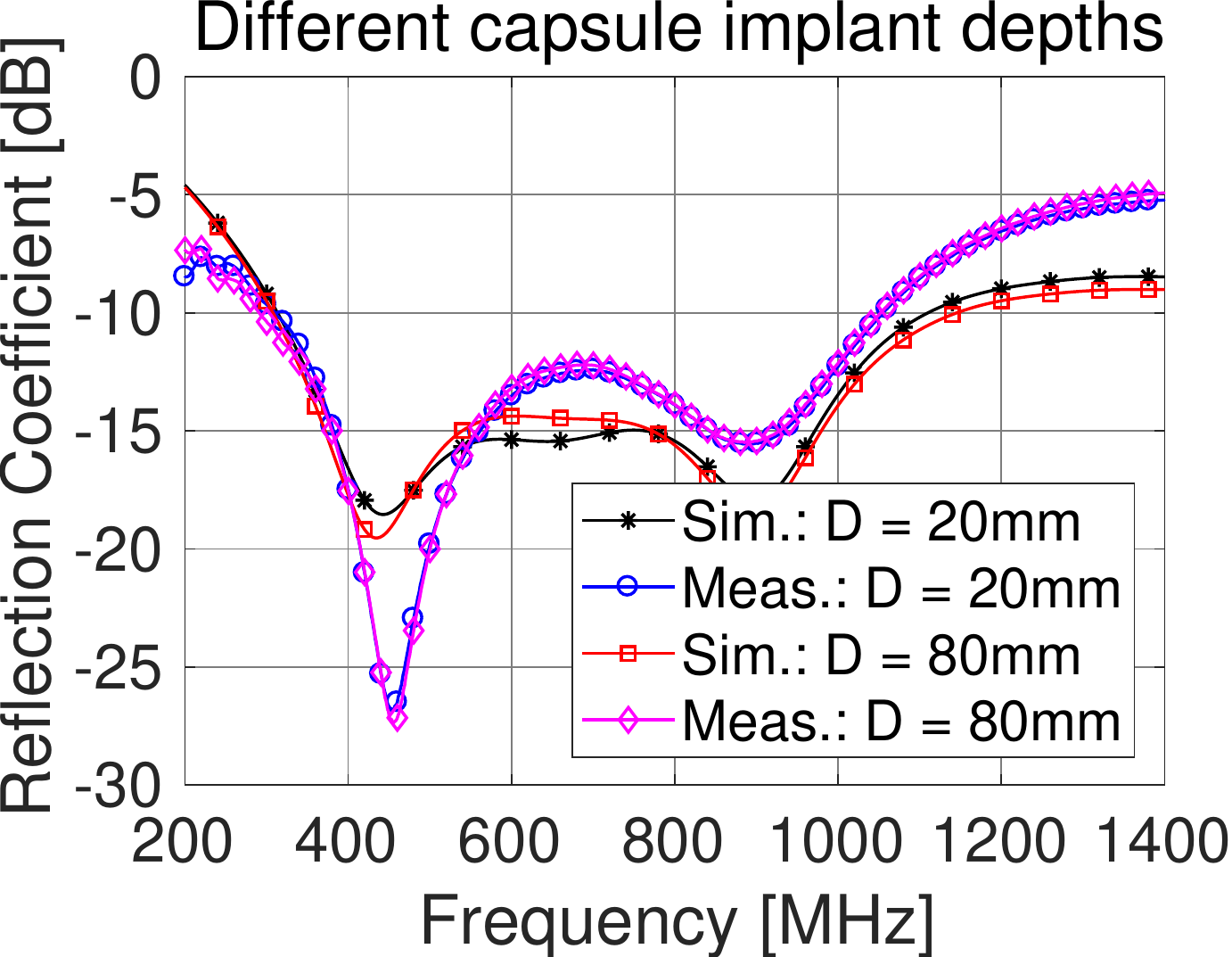}
				\label{fig:meas_in_body_implant_locations}}
			\subfigure[]{\includegraphics[scale=.30]{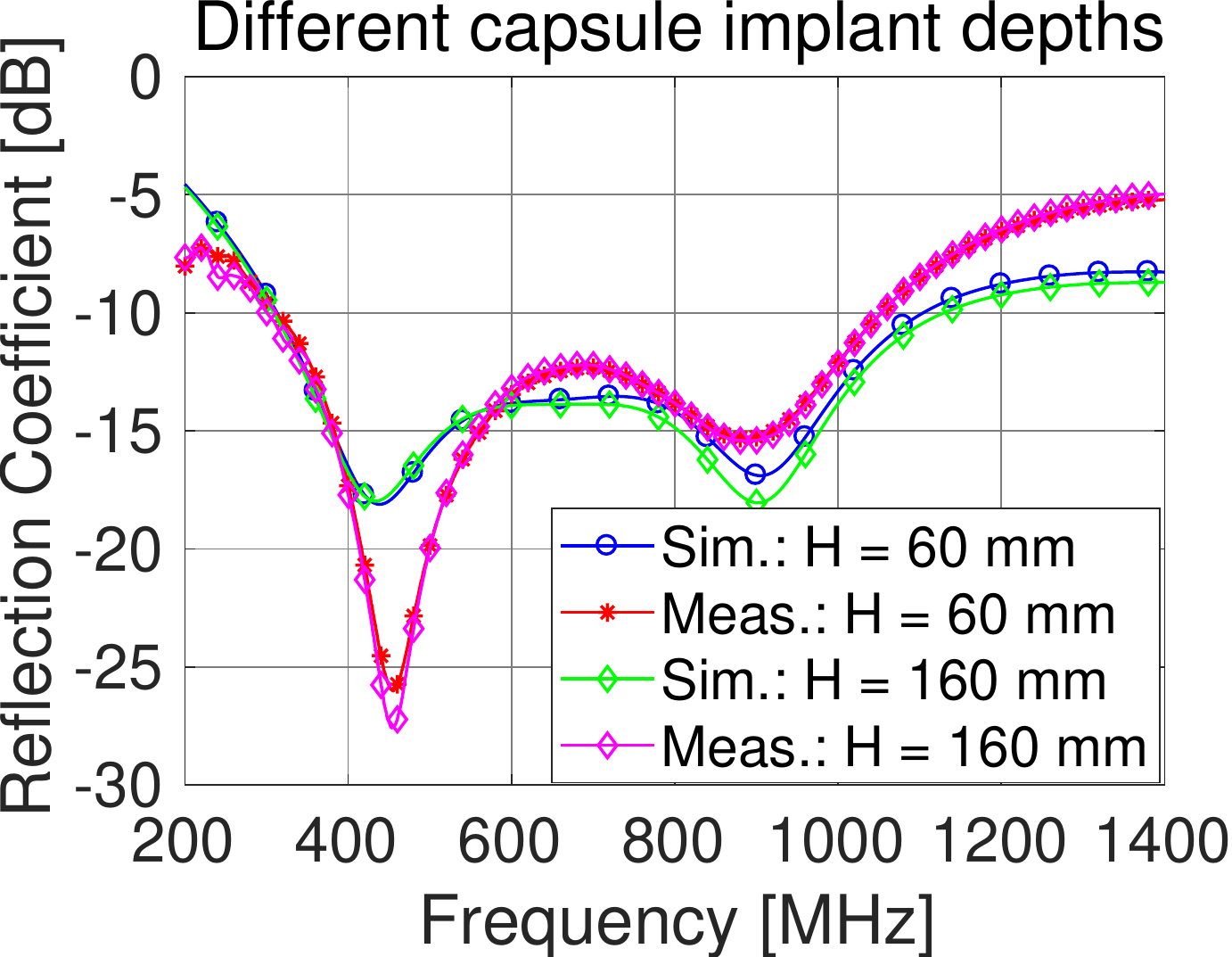}
				\label{fig:meas_in_body_implant_depth}}
			\caption{ Comparison of simulated and measured reflection coefficient: (a) $Y$-oriented capsule antenna at the center of the phantom; the simulated curve is identical to the one in Fig.~\ref{fig:sim_in_body_antenna_at_center}. The rest of the figures represent simulated and measured reflection coefficient for different (b) orientations, (c) locations, and (d) depth of the capsule in the phantom; the simulation results are taken from Fig.~\ref{fig:sim_in_body_implant_orientations}, Fig.~\ref{fig:depth_z} and Fig.~\ref{fig:depth_side}, respectively.}
			\vspace{-25pt}
			\label{fig:measured_ref_coeff_capsule_antenna}
		\end{center}
	\end{figure}

The comparison of the simulated and measured $|S_{11}|$ of a $Y$-oriented capsule antenna at the center of the liquid phantom is presented in Fig.~\ref{fig:sim_measured_antenna_at_center}. The simulated plot is identical to the one in Fig.~\ref{fig:sim_in_body_antenna_at_center}. The results show that the measured plot matches simulated one quite well. The lower resonance is slightly shifted upwards to $455$ MHz, while the second resonance is at $905$ MHz as in the simulations. The measured matching at $433$ MHz is $5$ dB better than in the simulations. This might come from the fabrication process of the prototype. Since the fabricated antenna was attached on the outer-wall of the capsule, there might be small air gaps between the antenna and capsule wall. The measured $|S_{11}|$ of the proposed capsule antenna is better than $-10$ dB across the frequency range from $280$ MHz to $1057$ MHz, which agrees well with the simulated matching bandwidth. 
	%\vspace{-15pt}
	
Several more measurements were performed with the orientations, depths, and locations of the capsule in the phantom as discussed in Section~\ref{sec:capsule_implementation_and_operating_environments}.  Fig.~\ref{fig:meas_in_body_implant_orientations} presents the comparison of simulated and measured $|S_{11}|$ of the antenna for three orientations of the capsule at the center of the phantom. Though slight differences in the resonance behavior are evident, the results show the validity of the antenna design and experimental evaluation process. 
Fig.~\ref{fig:meas_in_body_implant_locations} compares simulated and measured $|S_{11}|$ for different capsule locations $H$ from the side wall of the phantom, while maintaining capsule depth $D$ = $50$~mm. Fig.~\ref{fig:meas_in_body_implant_depth} presents $|S_{11}|$ for different capsule depths $D$ from the top with $H$ = $117.5$~mm. The measurements support the findings from simulations, i.e., the insignificant impact of the location and depth on the antenna resonance. The measured $|S_{11}|$ is better than $-10$~dB for the entire bandwidth of interest from $400$ to $600$~MHz regardless of the capsule locations and depths. Table IV shows a comparison of reported capsule antennas in the literature, designed frequencies are up to $1400$ MHz. The matching bandwidth of the proposed capsule antenna exceeds the existing solutions, while the realized peak gain is comparable. It should be noted that for an accurate comparison of the realized gain, all solutions need to be investigated in the same set-up using a phantom with identical geometry and dielectric properties.
\vspace{-12pt}
	\begin{table*}[ht]
		{\caption{ A comparison of reported capsule antennas in the literature, designed frequencies are up to $1400$ MHz: Antenna Type, operating frequency, Impedance matching bandwidth, radiation performance in the reported phantom}
			\label{table:Capsule_antenna_comparitive_table2}}
		\centering
		\begin{tabular}{p{0.08\linewidth}p{0.22\linewidth}p{0.07\linewidth}p{0.06\linewidth}p{0.07\linewidth}p{0.35\linewidth}}
			\hline
		 
			Ref. & Antenna type & Frequency (MHz) & Bandwidth (MHz) & Gain (dBi) & Phantom: tissue, shape, size (mm)\\
			\hline
			\cite{E_hatmi}  & Embedded, loop & 315 & 2 &&$\varepsilon_{\rm r} = 58, \sigma = 0.77$ S/m, cube,  640$^3$  \\
			\cite{E_FMerliTAP}  & Embedded, multilayer PIFA & 403 & 9 & $-29$ & Muscle, cyl., $\O$80 $\times$ 110\\
			\cite{E_Huang_letter}& Embedded, microstrip & 915 & 11 &  & Human body phantom \\
			\cite{E_SI_kwak2}  & Embedded, helical & 402 & 32 &  & $\varepsilon_{\rm r} = 56, \sigma = 0.8$ S/m, cyl.,$\O$75 \\
			\cite{E_SiKwak}  & Embedded, spiral & 402 & 70 &  & $\varepsilon_{\rm r} = 56, \sigma = 0.83$ S/m, cyl.,  $\O$150 $\times$ 150 \\
			\cite{Em_Lee_Yoo1}  & Embedded, fat-arm spiral & 450 & 75 &  & $\varepsilon_{\rm r} = 56, \sigma = 0.83$ S/m, cyl.  $\O$150 $\times$ 150 \\	
			%eSangLee  & Embedded, spiral & 402 & 90 & $\:-$ & Body, cyl.,  $\O$150 $\times$ 150 \\
			\cite{E_lee_dual_spiral}  & Embedded, dual-spiral & 402 & 98 &  & $\varepsilon_{\rm r} = 56, \sigma = 0.83$ S/m, cyl.,  $\O$150 $\times$ 150 \\
			\cite{E_sang_lee_conical}  & Embedded, conical spiral & 450 & 101 & & $\varepsilon_{\rm r} = 56, \sigma = 0.83$ S/m, cyl.,  $\O$150 $\times$ 150 \\
			\cite{E_sang_lee_Tbio} & Embedded, spiral & 500 & 104 & $-19.9$ & $\varepsilon_{\rm r} = 56.9, \sigma = 0.97$ S/m, cyl.,  $\O$150$\times$ 150\\
			\cite{E_sanglee_189}  & Embedded, dual-spiral & 402 & 189 &  & $\varepsilon_{\rm r} = 56, \sigma = 0.83$ S/m, cyl.,  $\O$150 \\
			%\hline	
			%\cite{Xiang_Senior_2_4} & Conformal, outer-wall patch with CSRR & 2400 & 10 & -5.2 & Damped sponge\\
			\cite{c_denys_TAP2017}& Conformal,~inner-wall microstrip & 434 & 17 & $-22$ & Muscle, sphere, $\O$100\\
			\cite{c_JFaerber_Tcir_2017}& Conformal, outer-wall helix & 433 & 20 & $-40.9$ & Muscle, cube, 190$^3$\\
			%cxichengawpl2011 & Conformal, patch & 2400 & 20 & -5.2 & Damped sponge\\
			\cite{Kiourti_P} & Conformal, outer-wall microstrip& 402 & 39.9 & $-29.6$ & Muscle, cube, 100$^3$ \\
			\cite{c_xiang_L} & Conformal, microstrip & 433 & 50 & $-9.6$ & $\varepsilon_{\rm r} = 58.1, \sigma = 0.83$ S/m, cyl., radius and height are not reported \\
			\cite{pathloss_3_Yann_mahe} & Conformal, microstrip& 434 & 53 & $-33$ & $\varepsilon_{\rm r} = 49.6, \sigma = 0.51$, cyl., $\O$ 200 \\
			\cite{arefin} & Conformal, inner-wall patch& 433 & 124.4 & $-36.9$ & $\varepsilon_{\rm r} = 56.4, \sigma = 0.82$, cyl. $\O$ 100 $\times$ 200 \\
			\cite{c_lijie_AWPL} & Conformal, dipole & 402 & 158 & $-37$ &Skin, cube, 180$^3$\\
			\cite{c_KKwon} & Conformal, PIFA & 1400 & 150 & $-29.7$ & Human body model\\
			\cite{c_rupom_das_TAP2017} & Conformal, inner-wall loop & 915 & 185 & $-19.4$ & $\varepsilon_{\rm r} = 55, \sigma = 0.95$ S/m, cube, 100$^3$\\
			\cite{Izdebski_TAP} & Conformal, outer-wall dipole & 1400 & 200 & $-26$ &Muscle, box, 350 $\times$ 350 $\times$ 200\\
			\cite{c_rajagopalan} & Conformal, inner-wall dipole & 1400 & 200 & $-36$ &Muscle, box, 600 $\times$ 300 $\times$ 400\\
		%	& & 	& 300 &-26&Small intestine, human body phantom\\
			\cite{Sumin_Kim_TAP_L} & Conformal, outer-wall loop & 500 & 260 &  & $\varepsilon_{\rm r} = 56.4, \sigma = 0.82$, cyl. $\O$150\\
			%\cite{c_vivek} & Conformal, spiral & 402 & 300 & $-32$ & $\varepsilon_{\rm r} = 58, \sigma = 0.77$ S/m,\\
		
			\cite{Jing_Lim} & Conformal, ouer-wall PIFA & 500 & 562 & $-25.2$ & $\varepsilon_{\rm r} = 56, \sigma = 0.83$ S/m, box, 200 $\times$ 150 $\times$ 100 \\
			\cite{Rula_huang} & Conformal, outer-wall loop & 500 & 785 &  & $\varepsilon_{\rm r} = 56.4, \sigma = 0.82$, box, 200 $\times$ 150 $\times$ 100 \\
			This paper & Conformal, outer-wall loop & 433 & 795 & $-35$ & Colon, box, 235 $\times$ 220 $\times$ 100\\
			This paper & Conformal, outer-wall loop & 433 & 795 & $-23.5$ & Human body phantom\\

			\hline
		\end{tabular}
	\end{table*}
	\subsection{On-Body Antenna}
	\label{sec:measurements_on_body_antenna}
\begin{figure}[t]
		\begin{center}
			\subfigure[]{\includegraphics[scale=.2]{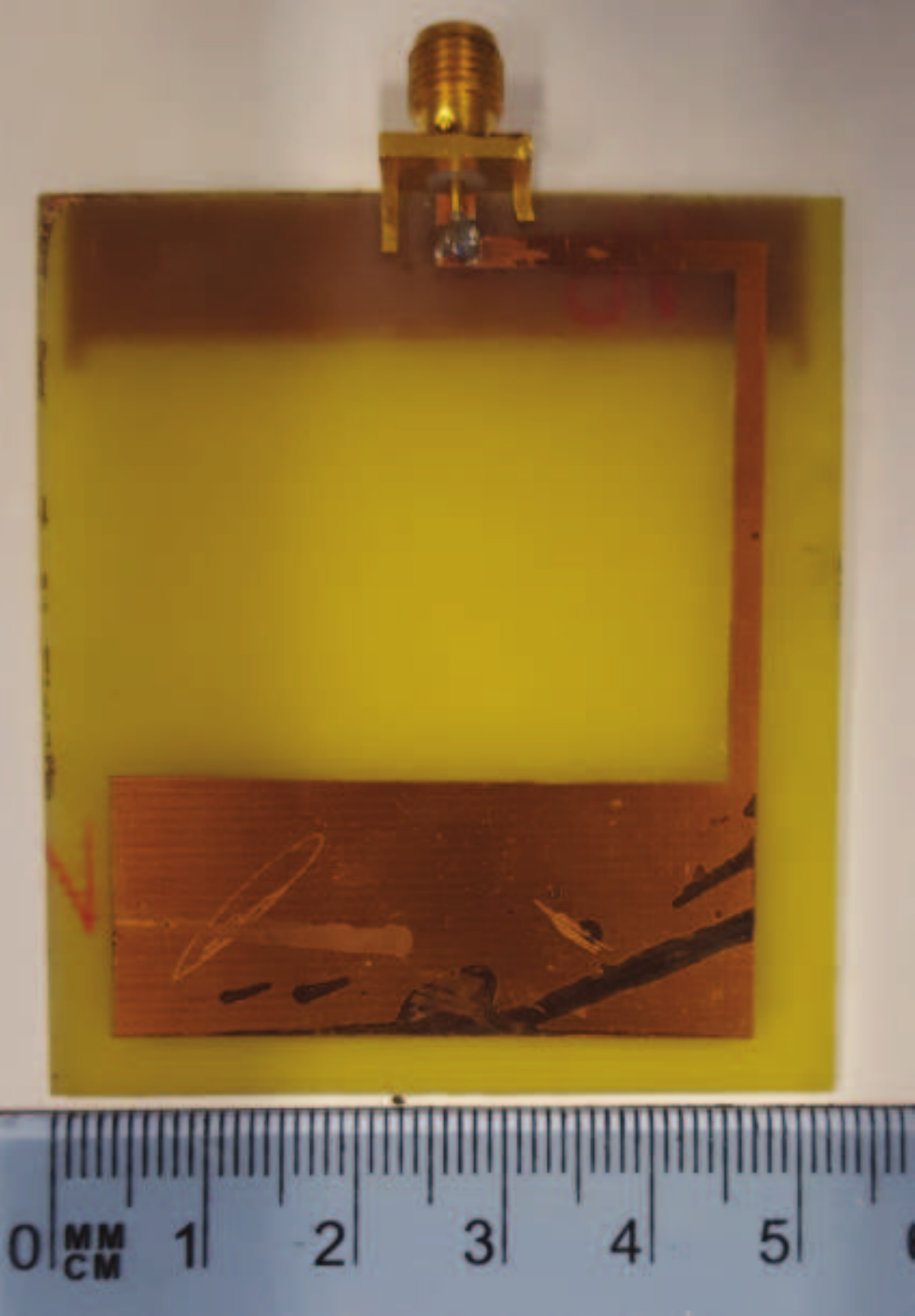}
				\label{fig:fabricated_on_body_antenna_top}}
			\subfigure[]{\includegraphics[scale=.172]{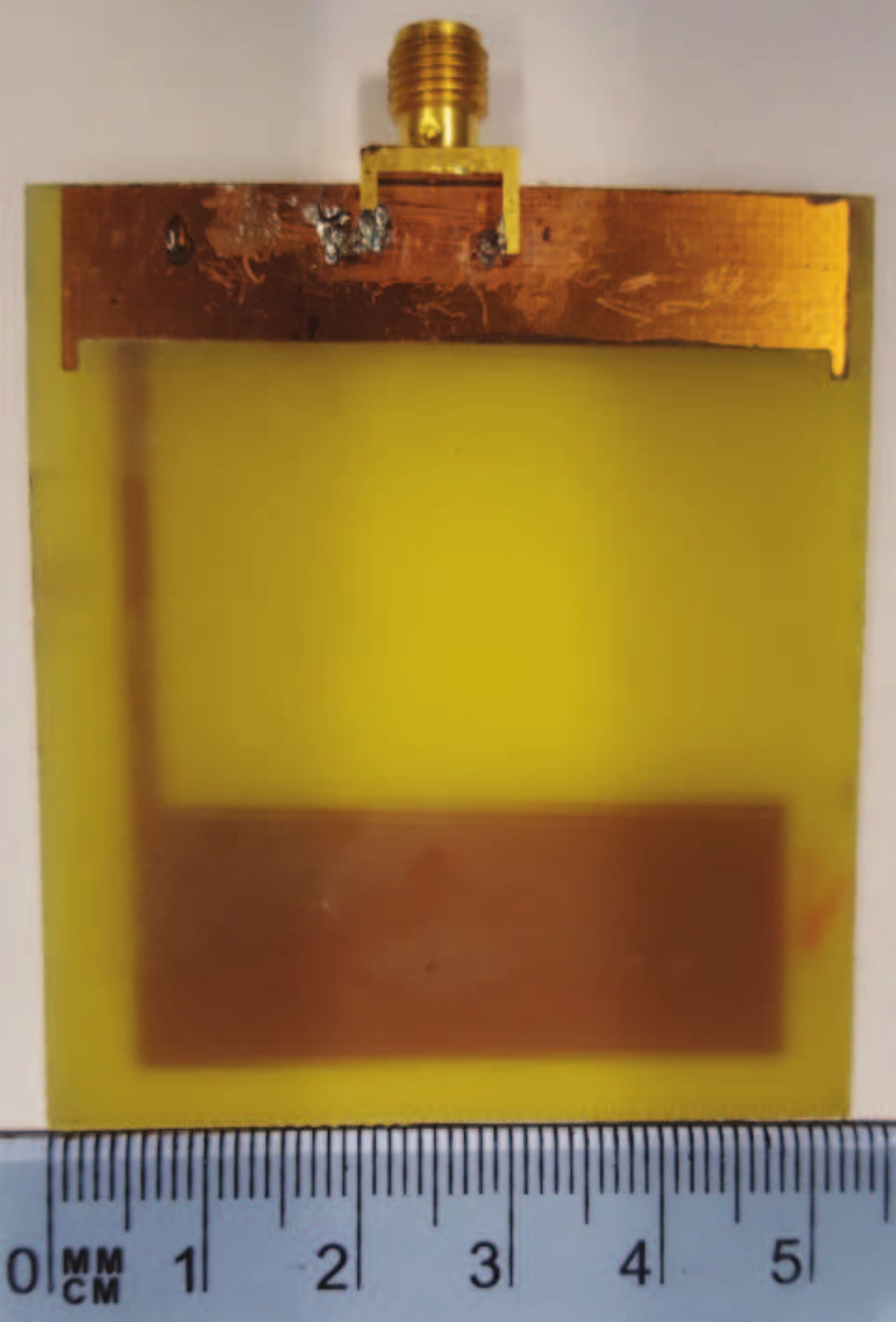}
				\label{fig:fabricated_on_body_antenna_back}}
			\caption{Fabricated on-body antenna: (a) top view, and (b) bottom view. }
			\vspace{-15 pt}
			\label{fig:fabricated_on_body_antenna_prototype}
		\end{center}
	\end{figure}

\begin{figure}[t!]
		\begin{center}
			\resizebox{0.6\hsize}{!}{\includegraphics[scale=.4]{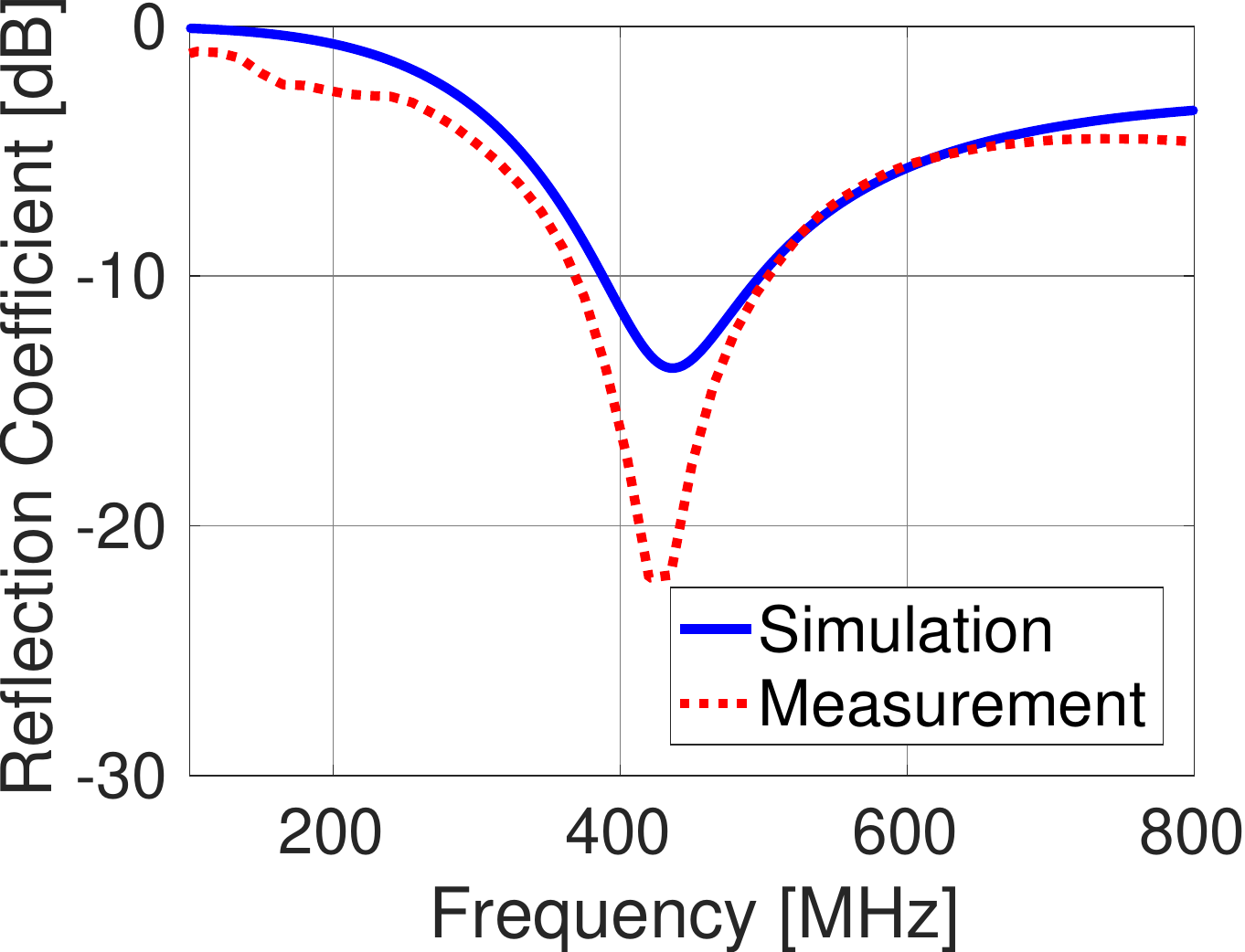}}
			\caption{Simulated and measured reflection coefficient of the on-body antenna on the liquid phantom; the simulation curve is taken from Fig.~\ref{fig:S11_on_body_simulation}.}
			\vspace{-10pt}
			\label{fig:meas_receiver_antenna_liquid}
		\end{center}
	\end{figure}

	The on-body antenna was fabricated on the same FR4 substrate as in the simulation, detailed in Section~\ref{sec:on_body_antenna_structure}. The fabricated on-body monopole antenna is shown in Fig.~\ref{fig:fabricated_on_body_antenna_prototype}. The \textit{in-vitro} measurement was performed using the liquid phantom, introduced in Section~\ref{sec:meas_phantom}, as shown in Fig.~\ref{fig:measurement_setup}. The on-body antenna, which was placed on the outer wall of the liquid container, was connected to the VNA port-2 through a coaxial cable, while VNA port-1 was open-ended during this measurement. The simulated and measured $|S_{11}|$ follow the same trend, as can be seen in Fig.~\ref{fig:meas_receiver_antenna_liquid}, although the maximum magnitude differs slightly. Still, the antenna resonates at the same frequency as in the simulation i.e., $420$~MHz, with $-10$~dB impedance bandwidth of $120$~MHz.  
	
%Fig.~\ref{fig:in_vivo_measurement_setup_bareskin} shows the \textit{ex-vivo} measurement set-up, with the simulated and measured $|S_{11}|$ in Fig.~\ref{fig:meas_receiver_antenna_bareskin}. The two curves agree well, with identical antenna resonance at $465$~MHz and the better matching than $-10$~dB at $433$~MHz.	

\begin{figure}[t!]
		\begin{center}
			\subfigure[]{\includegraphics[scale=.12]{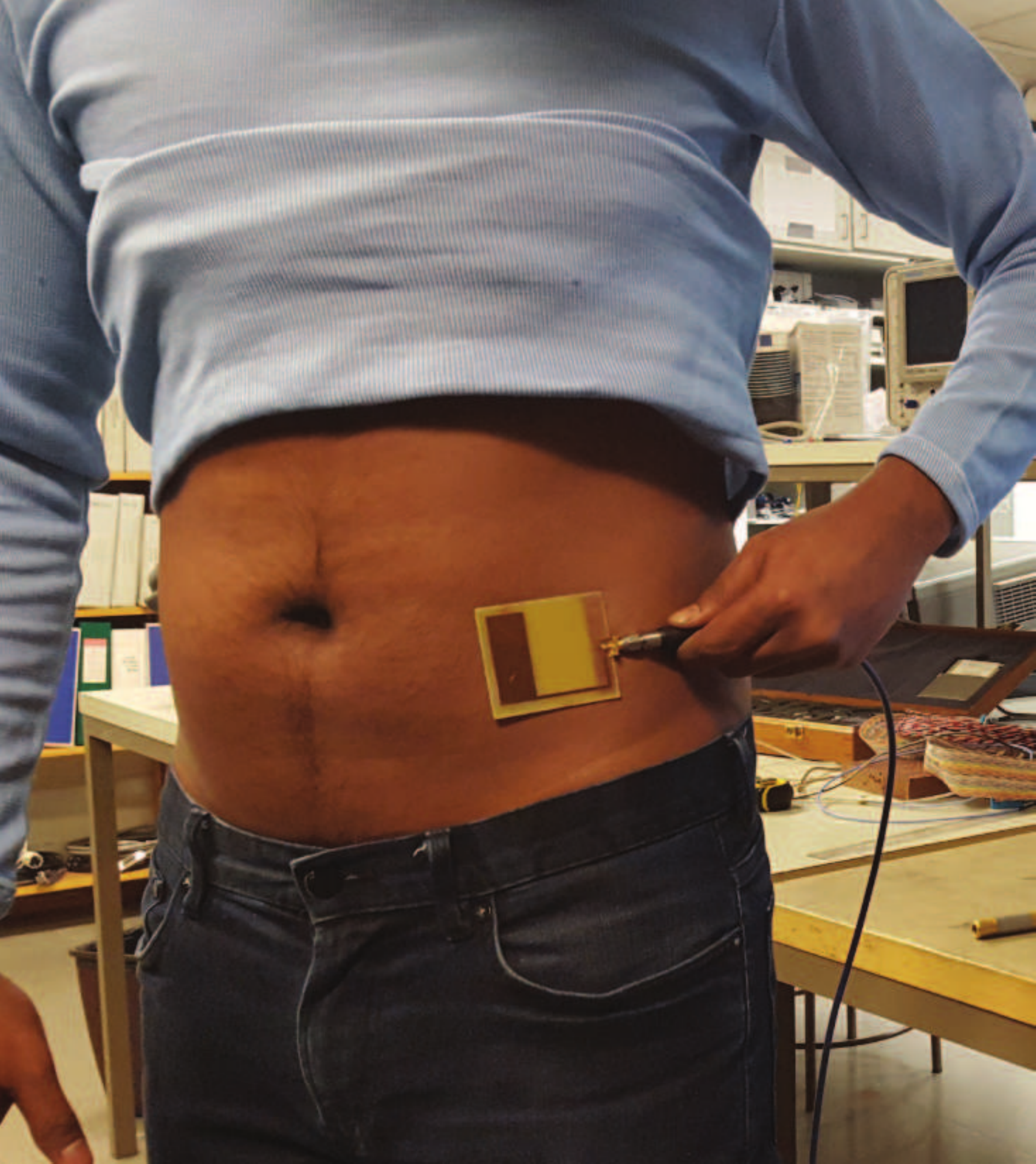}
				\label{fig:in_vivo_measurement_setup_bareskin}}
			\subfigure[]{\includegraphics[scale=.35]{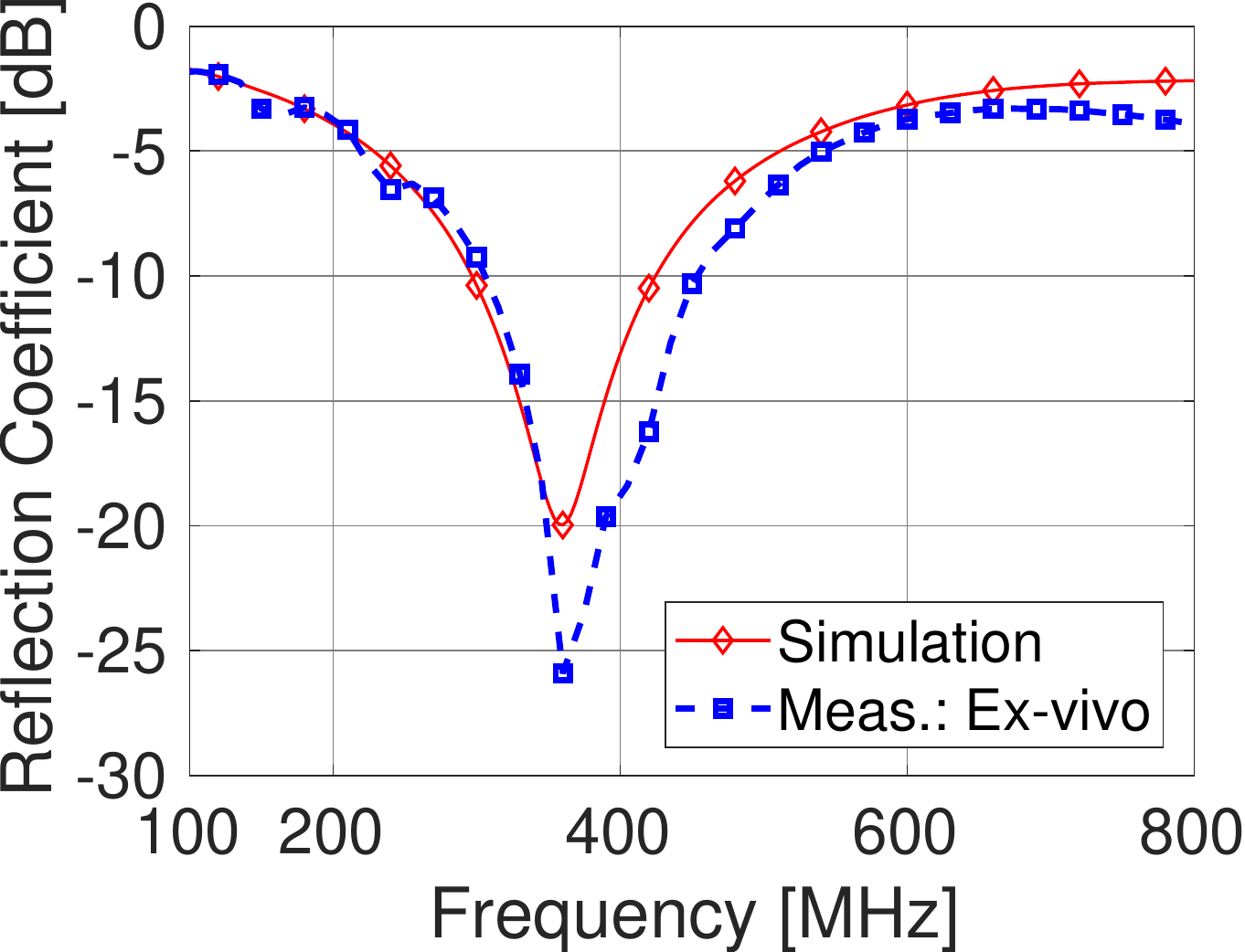}
				\label{fig:meas_receiver_antenna_bareskin}}
			\caption{(a) \textit{Ex-vivo} measurement set-up of the on-body antenna, and (b) Simulated and measured reflection coefficient of the antenna, where the simulation results are taken from Fig.~\ref{fig:S11_on_body_simulation}.}
			\vspace{-15pt}
			\label{fig:ex_vivo_measurement_setup_and_results_wo_cloth}
		\end{center}
	\end{figure}

In addition to measurements with the phantom, the antenna matching was studied with a real human body of BMI $22$ in a laboratory environment. In agreement with the simulations in Section~\ref{sec:on_body_antenna_multi_tissue_phantom} that use the realistic human body model, the \textit{ex-vivo} measurements include the on-body antenna placed on the abdominal skin directly. Fig.~\ref{fig:in_vivo_measurement_setup_bareskin} shows the \textit{ex-vivo} measurement set-up. Fig.~\ref{fig:meas_receiver_antenna_bareskin} presents the measured and simulated $|S_{11}|$ having a similar profile, though their best matching level differs. Nevertheless, the on-body antenna resonates at $360$~MHz as in the simulations with the CST Gustav voxel human body model. The comparison consistently shows better matching than $-10$ dB at $433$~MHz. 	
%\vspace{-10pt}
%	\begin{figure}[t!]
%		\begin{center}
%			\subfigure[]{\includegraphics[scale=.12]{fig/in_vivo_measurement_cloth2.eps}
%				\label{fig:in_vivo_measurement_setup_cloth}}
%			\subfigure[]{\includegraphics[scale=.33]{fig/meas_receiver_antenna_cloth.eps}
%				\label{fig:meas_receiver_antenna_cloth}}
%			\caption{(a) \textit{Ex-Vivo} matching measurement with cloth, i.e., set-up \#1, and (b) Comparison of simulated and measured reflection coefficient; the simulation curve is taken from Fig.~\ref{fig:S11_on_body_simulation}.}
%			\vspace{-15pt}
%			\label{fig:ex_vivo_measurement_setup_and_results_w_cloth}
%		\end{center}
%	\end{figure}
		
	\subsection{Path Loss}
	The propagation path loss between the capsule and on-body antenna was studied using the proposed antennas. The homogeneous colon tissue phantom model was considered both in simulations and measurements. A full-wave time domain simulation was conducted to estimate the propagation loss in the tissue. The path loss for a particular capsule and on-body antenna location is defined as the ratio of the input power at the transmit capsule antenna to the output power at the on-body antenna port. In terms of transmission coefficient of the two-port network, it is expressed as $PL|_{\rm dB} = -10\log_{10}|S_{21}|$.
	
	\begin{figure}[t!]
		\begin{center}
			\resizebox{0.90\hsize}{!}{\includegraphics[scale=.45]{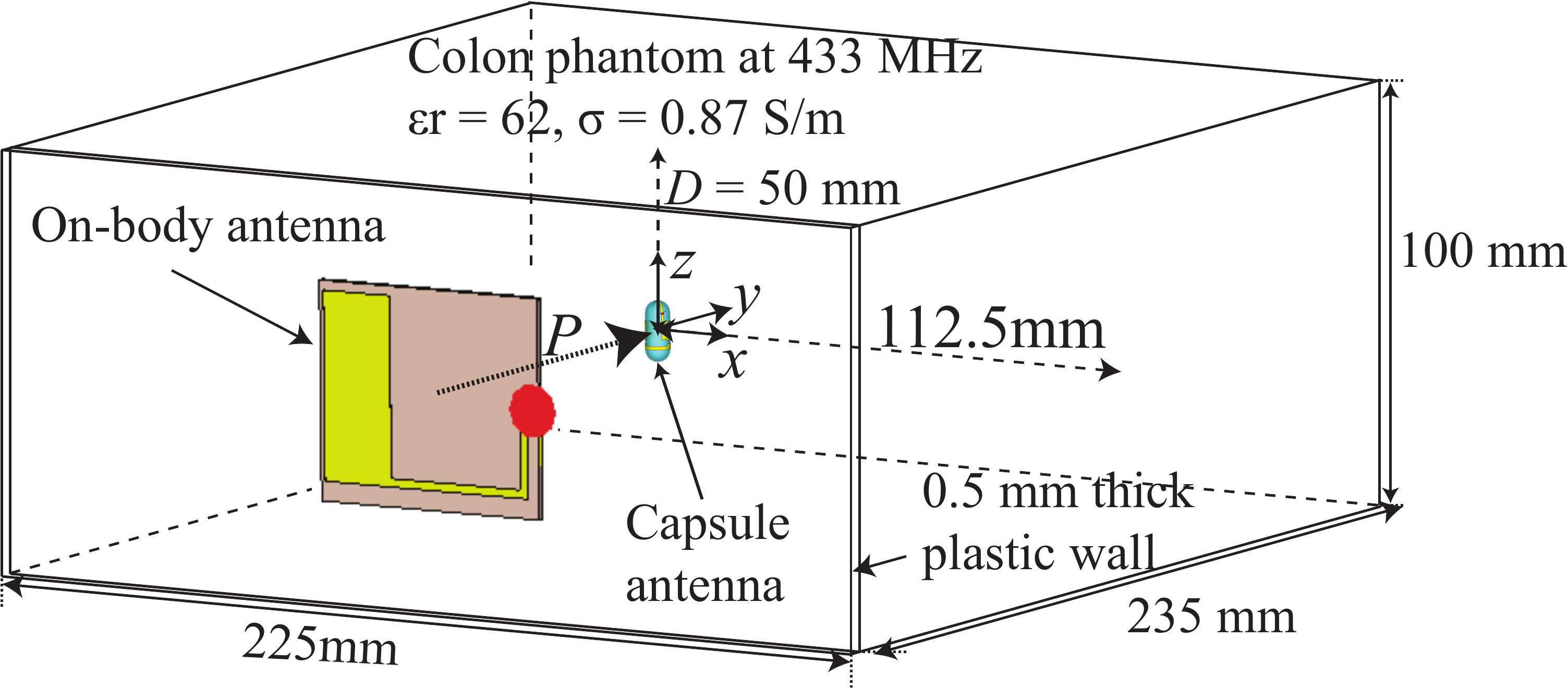}}
			\caption{Set-up for path loss simulation of the in-body to on-body radio link.}
			\vspace{-25pt}
			\label{fig:sim_arrangement_pathloss}
		\end{center}
	\end{figure}
\begin{figure}[t!]
	\begin{center}
	\includegraphics[scale=.6]{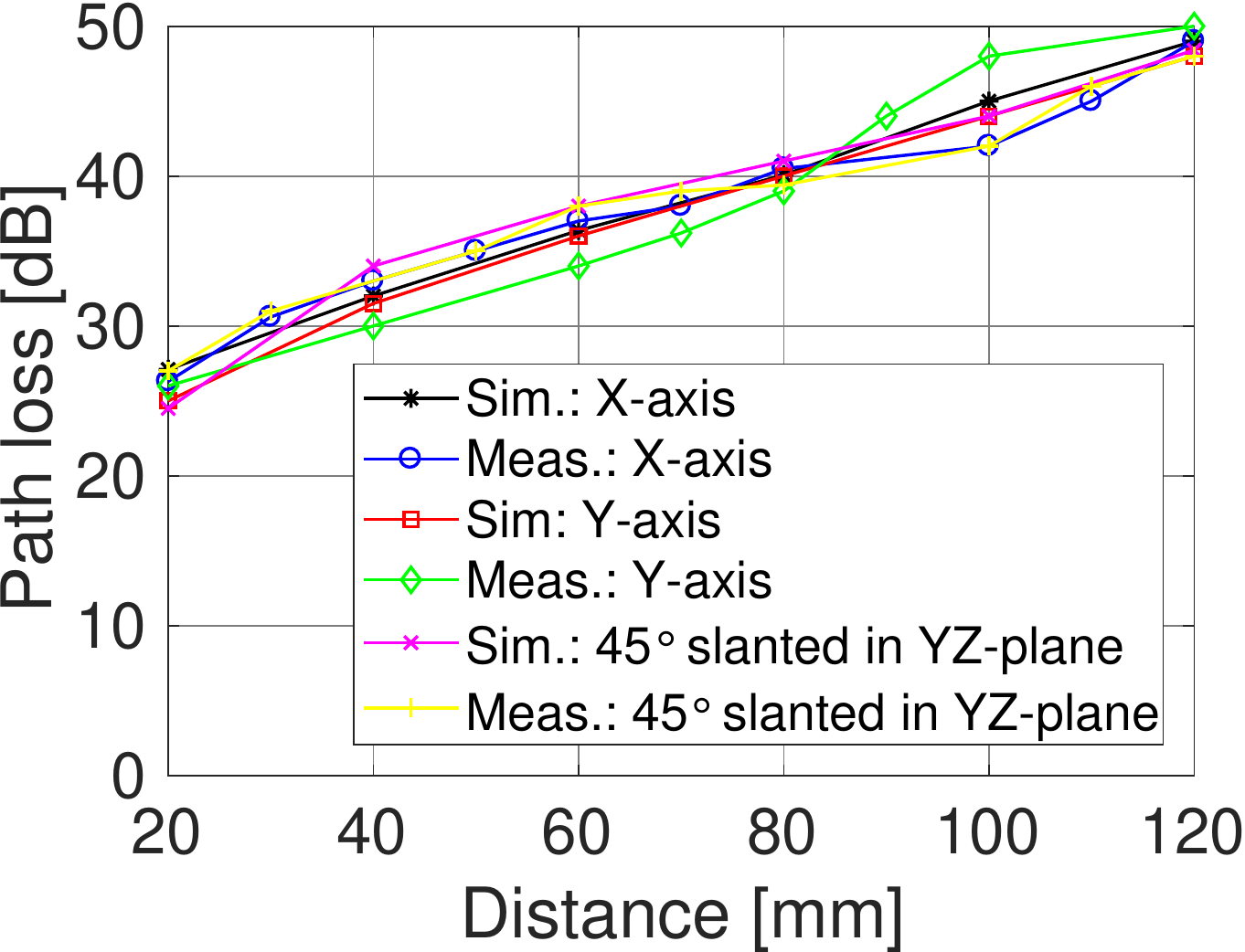}
	\caption{Propagation path loss between in-body capsule and on-body antennas.}
	\vspace{-25pt}
	\label{fig:path loss_sim_meas}
	\end{center}
	\end{figure}

	Fig.~\ref{fig:sim_arrangement_pathloss} illustrates the path loss simulation setup. Three different orientations of the capsule are considered, such that they are along the \textit{X}-axis, \textit{Y}-axis, and $45^\circ$ slanted on the \textit{YZ}-plane. For each orientation, the transmit-receive distance was changed by moving the capsule antenna along \textit{Y}-axis away from the receive antenna, i.e., $P$ increases, while maintaining the same $D = 50$~mm. The distance between the antennas includes $0.5$~mm thickness of phantom's plastic container and $1$~mm air gap in between the on-body antenna and phantom's wall. The path loss was estimated at six distances between $P = 20$ to $120$~mm for each of the above-mentioned capsule orientations. The corresponding \textit{in-vitro} measurements were performed using the liquid phantom introduced in Section~\ref{sec:meas_phantom} as shown in Fig.~\ref{fig:measurement_setup}. The capsule and on-body antennas were connected to the VNA ports 1 and 2, respectively. A thru-calibration of the VNA was performed in order to subtract the cables' frequency response. The path loss was obtained from readings of the scattering parameters on the VNA. The simulated and measured path loss are plotted together in Fig.~\ref{fig:path loss_sim_meas}, showing their close agreement for all the tested capsule orientations. The mean simulation path loss error is $-0.4$, $-0.4$ and $-0.3$~dB for the \textit{X}-, \textit{Y}- and \textit{YZ}-oriented capsule, respectively, whereas standard deviation is $1.3$, $2.1$ and $1.4$~dB. The minimum path loss is observed when capsule antenna is $45^\circ$ slanted on the \textit{YZ}-plane. The maximum path loss of $50$~dB is observed in the measurement when capsule antenna is \textit{Y}-oriented at the center of the phantom. The path loss differences across capsule orientations are mainly due to polarization mismatch between in- and on-body antennas. The simulated axial-ratio towards negative \textit{Y}-axis direction at $433$ MHz of the \textit{YZ}-, \textit{X}- and \textit{Y}-oriented capsule at the center of the colon-tissue phantom is $8.4$, $9$ and $13$ dB, respectively. The values indicate elliptical to linear polarization of the transmitted far-fields. The path loss of the \textit{Y}-oriented capsule is slightly higher most likely due to more a pronounced polarization mismatch than other orientations. The variations in polarization mismatch between in- to on-body antennas can be mitigated by using a circularly polarized antenna at one end of the link, or a dual-polarized antenna at the receiver side. However, the proposed antenna solutions do not solve the polarization mismatch issue, but future works include the design of circularly polarized antennas. According to \cite{ken_ichi_pathloss}, the acceptable path loss for an improved WCE was $73.3$~dB\footnote{ The derivation of the tolerable path loss estimate assumes, for example, the channel bandwidth of $2.4$ MHz, $0$ dBm input power to the capsule antenna, fading margin of $10$ dB, channel signal-to-noise ratio of $16$ dB, and the noise figure of $10$ dB. Then the link is capable of transferring $2.0$ Mbps, which suffices for sending $6$ camera images per second.}. The proposed antenna solutions, therefore, are capable of supporting the improved WCE.
	\vspace{-10pt}
	
	%\footnote{Since the input power given in Effective Isotropic Radiated Power (EIRP) is set to $-16$~dBm in \cite{ken_ichi_pathloss}, the tolerable path loss estimate assumes $0$~dBm input power to the capcule antenna and path loss calculation in our study assumes antenna gain of $-16$~dBi.} for an improved WCE. The proposed antenna solutions therefore are capable of supporting the improved WCE.    
	
%	\begin{table}
%	\begin{center}
%	{\caption{A summary of the mean and standard deviation of the simulation path loss error for each capsule orientation [db]} 
%		\label{table:std_deviation}}
%	\begin{tabular}{cccc} \hline
%	& $X$-axis  & $Z$-axis  & $YZ$-axis \\
%	&  orientated  & orientated & orientated\\
%	\hline
%	Mean & -0.4 & -0.4 & -0.3\\
%	Standard deviation & 1.3 & 2.1 & 1.38\\
%	\hline
%	\end{tabular}
%	\end{center}
%	\vspace{-10pt}
%	\end{table}

	\section{Conclusion}

In this paper, we present novel small antenna solutions of an improved WCE operating at $433$ MHz ISM band. An ultrawideband conformal loop antenna is proposed for the in-body transmitting capsule, while the on-body receiving antenna is a printed monopole with a partial ground plane. The in-body antenna utilizes the outer-wall of the capsule and leaves the inner space for the other electronic components. For a required impedance matching of $-10$~dB, the proposed antenna supports a bandwidth of $795$~MHz, from $309$ to $1104$ MHz. The ultrawideband matching enables the capsule antenna to overcome the detuning effects due to orientations and locations of the capsule, electronics modules in the capsule and proximity of the capsule to various different tissues in GI tract. The on-body antenna was numerically evaluated on the colon-tissue phantom and a realistic human body model. \textit{In-vitro} measurements on a liquid phantom and \textit{ex-vivo} measurements on a real human body were performed for validations. The measured $-10$ dB impedance matching from $390$ MHz to $500$ MHz shows good agreement with simulated results. Finally, the path loss for the radio link between an in-body capsule transmitter and an on-body receiver was studied using the proposed antenna solutions. The path loss is less than $50$ dB for all the capsule orientations and locations in the body, opening the way for improved ingestible WCE by supporting higher-data-rate radio links.

	%\vspace{7pt}
%	\section*{Acknowledgment}
%	The authors would like to acknowledge Dr. \ Vasilii Semkin, Dr.\ Ilya Anoshkin for their support in antenna fabrication, and Premix Oy for providing the substrate material.

	\bibliographystyle{IEEEtran}% bib style
	\bstctlcite{IEEEexample:BSTcontrol}
	\bibliography{ref}% your bib database
	
	% that's all folks
	\end{document}